\documentclass[preprint]{imsart}

\usepackage{amsthm,amsmath,natbib}
\RequirePackage[colorlinks,citecolor=blue,urlcolor=blue]{hyperref}
\usepackage{graphicx}
\startlocaldefs
\endlocaldefs

\begin{document}

\begin{frontmatter}

\title{Design-adherent estimators for network surveys}
\runtitle{Adherent estimators}


\author{\fnms{Steve}
  \snm{Thompson}\corref{}\ead[label=e1]{thompson@sfu.ca}}
\address{Department of Statistics and Actuarial Science\\8888
  University Drive\\Burnaby, BC V5A 1S6 Canada\\\printead{e1}}
\affiliation{Simon Fraser University}

\runauthor{Steve Thompson}

\begin{abstract}
  Network surveys of key populations at risk for HIV are an essential
  part of the effort to understand how the epidemic spreads and how it
  can be prevented.  Estimation of population values from
  the sample data has been probematical, however, because the
  link-tracing of the network surveys includes different people in the
  sample with unequal probabilities, and these inclusion probabilities
  have to be estimated accurately to avoid large biases in survey
  estimates.  A new approach to estimation is introduced here, based
  on resampling the sample network many times using a design that
  adheres to main features of the design used in the field.  These
  features include network link tracing, branching, and
  without-replacement sampling.  The frequency that a person is
  included in the resamples is used to estimate the inclusion
  probability for each person in the original sample, and these
  estimates of inclusion probabilities are used in an
  unequal-probability estimator.  In simulations using a population of
  drug users, sex workers, and their partners for which the actual
  values of population characteristics are known, the design-adherent
  estimation approach increases the accuracy of estimates of
  population quantities, largely by eliminating most of the biases.
\end{abstract}


\begin{keyword}
\kwd{Network sampling}
\kwd{Snowball sampling}
\kwd{Respondent-driven sampling}
\end{keyword}

\end{frontmatter}

\section{Introduction}

\begin{figure}
\centering
\includegraphics[width=1.0\linewidth]{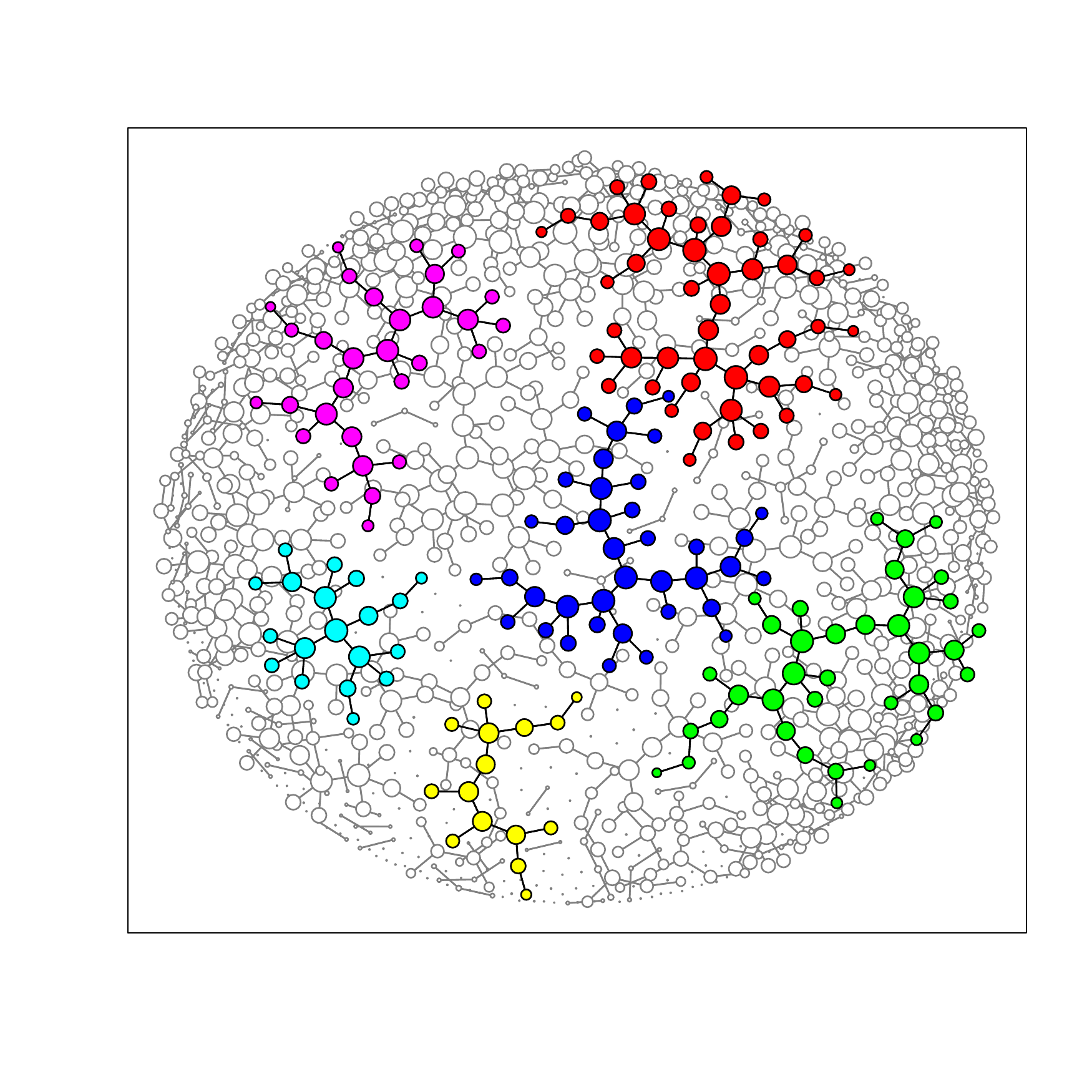}
\caption{A network sample of 1200 people from a high-risk hidden
  population. Some sample components are highlighted.  The network
  sample is resampled many times using a network sampling design that
  adheres closely to the original survey design.  The frequency that a
  person is included in the resamples (circle diameter) is used to estimate the person's
  inclusion probability in the original sample.  These estimated
  inclusion probabilities enable estimates of hidden-population
  characteristics.  }
\label{fig:sample}
\end{figure}

Network sample surveys have become an essential part of the
effort to understand and alleviate the HIV epidemic worldwide.  Much
of the spread of the epidemic is over sexual and drug-using
connections among hidden or hard-to-access populations.  Network
surveys that follow social links between members of the hidden
population are effective and often the only feasible way to reach into
the key populations most at risk for HIV.  Estimation of the
characteristics of the key population from the survey data has been
problematical because the link-tracing designs find different members
of the population with unequal probabilities depending on network
connection patterns.  The core problem of network sampling is to
estimate the inclusion probabilities of the actual design accurately.

A new approach to estimation for network surveys
is described in this report.  The idea underlying the method is
to select many resamples from the sample network using a design that
adheres closely to features of the actual design used in selecting the
sample from the hidden population.  For each person in the sample, the
inclusion frequency of that person in the resamples is used as an
estimate of that person's relative inclusion probability in the
original sample.  The estimated inclusion probabilities are then used
to estimate population characteristics and calculate confidence
intervals.

 Figure \ref{fig:sample} shows a sample of 1200 people selected by
 Respondent-Driven Sampling (RDS) in which each respondent has been
 given up to 3 coupons to recruit some of his or her partners into the sample.
  Several sample
components are highlighted.  Each of these components was started from
a single seed.  Links were traced to add additional people to the
sample.  A person could not be recruited more than once so the
sampling is without-replacement.

If instead we used a Snowball (SB) design
allowing unlimited recruitments or up to some high maximum like 15, the
sample would tend to have larger components and much more branching.
With either of these designs, the sample is a network tree, with only a
single path from one individual to another that does no backtracking.
Each individual may have more links than are found in the sampling,
because of the limits on recruitment and the chances that a recruiting
attempt is successful or not.  

In some surveys (termed RDS+ or SB+ here), extra
social network interviews or relationship matching determine 
additional links between sample members.  So a person in the (red)
component at the top of the figure might have an additional link, that
was not used in recruiting, to another person in that or another
component.  With the extra links the sample, like the population,
would not be a tree but a more general network structure containing
cycles or multiple paths between individuals.

A random walk design, which allows no branching, would produce a
sample very different from Figure \ref{fig:sample}.  Each respondent would
be given exactly one coupon, to select up to one partner, and that
partner would in turn be given one coupon, and so on.  The sample
components would look like crooked stems, rather than trees.  Random
walk designs are seldom if ever used in actual network surveys. But
the theoretical inclusion probabilities of a with-replacement random
walk is used as the basis for currently used estimators.  The
discrepancy between the assumed theoretical design and the actual
survey design used gives rise to the statistical problems of the
currently  used estimators.

The branching, without-replacement designs have several advantages
over a non-branching or with-replacement design.  The
without-replacement sampling means that each selection is a new node,
not already selected, and  so the number of distinct units in
the sample grows steadily.  The branching allows the sample to spread
and grow rapidly when a link-rich, highly connected area of the social
network is encountered.

An under-appreciated feature of the branching, without-replacement
network sampling designs such as RDS and SB designs is their ability
to self-allocate, so that proportionally more sampling effort is made
in the highly connected parts of the hidden population.  Suppose that
in the hidden population are two separate components, so that no path
of links leads from one to the other, and that one of the components
is more densely connected, having more links and paths than the other.
If an equal number of random-walk seeds are placed in each component,
after any number of steps there will still be the equal numbers of
sample nodes selected in each component.  With a branching design, if
equal numbers of seeds are selected in each component, there will
after several steps be many more sample nodes in the component that
has the more links and connections, because at any time there are more
links out from the sample that can be followed, so probabilistically
more of the selections will be made in the more densely-connected
component.  This will happen naturally even when the recruiting of new
sample members is done by the hidden-population respondents.  This
self-allocation with the branching, without-replacement design serves
also to make the designs effective at finding densely connected
centers in the network, and once the sample finds those the
recruitments spread outwork in all directions from those centers.

\subsection*{How the estimation method works}

In traditional survey sampling with unequal probabilities of inclusion
for different people, typical estimators divide an observed value
$y_i$ for the $i$th person by the inclusion probability $\pi_i$ that
person.  The variable of interest $y_i$ might be 1 if the
person tests positive for a virus and 0 otherwise, or might be the
number of partners that person reports having.  The inverse-weighting
gives an unbiased or low-bias estimate of the population proportion or
mean of that variable.  In network surveys the
inclusion probabilities are unknown so they need to be estimated.

The estimators described in this report estimate the inclusion probability of
each person in the sample by selecting many resamples from the
network sample data using a design that adheres in key features to the
actual survey design used to find the sample.    In particular, the resampling
design is a link-tracing design done without-replacement and with
branching, as was the original design.  The frequency $f_i$ with
which an individual is included in the resamples is used as an
estimate of that person's inclusion probability $\pi_i$.  

What we do is select $T$ resamples $S_1, S_2, ..., S_T$ from the
sample network data.  There are two aproaches to selecting the
sequence of resamples.  In the repeated-samples approach each resample
is selected independently from seeds and progresses step-by-step to
target resample size independently of every other resample, so we get
a collection of independent resamples.  in the sampling-process
approach each resample $S_t$ is selected from the resample $S_{t-1}$
just before it by randomly tracing a few links out and randomly
removing a few nodes from the previous resample and using a small rate
of re-seeding so we do not get locked out of any component by chance.
It is this Markov resampling process approach that we use for the
simulations in this paper because it is so computationally efficient.

For an individual $i$ in the original sample, there is a sequence of
zeros and ones $Z_{i1}, Z_{i2}, ..., Z_{iT}$, where $Z_t$ is 1 if that
person is included in resample $S_t$ and is 0 if the person is not
included in that resample.  The inclusion frequency for person $i$ is 
\begin{equation}
f_i =\frac{1}{T}\left( Z_{i1} + Z_{i2} + ... + Z_{iT} \right)
\label{eq:fi}
\end{equation}

In Figure \ref{fig:sample} the circle representing individual $i$ in
the sample is drawn with diameter proportional to the estimated
inclusion probability $f_i$ of that individual.  Individuals centrally
located in sample components tend to have high values of $f_i$.  That
is because there are more paths, and paths of higher probability,
leading the sample to those individuals.  Also, individuals in larger
components tend to have larger $f_i$ than individuals in smaller
components, so that the method is estimating inclusion probability of
an individual relative to all other sample units, not just those in
the same component or local area or the sample This is because of the
self-allocation of the branching design, even in the absence of
re-seeding, to areas of the social network having more links and
connected paths.

The estimator of the mean of a characteristic $y$ in the hidden
population is then 
\begin{equation}
\hat\mu_f = \frac{\sum (y_i/f_i)}{\sum (1/f_i)}
\label{eq:est}
\end{equation}
where each sum is over all the people in the sample.  If the actual
inclusion probabilities $\pi_i$ were known and replaced the $f_i$ in
Equation \ref{eq:est} we would have the generalized unequal
probability estimator $\hat\mu_\pi$ of Brewer \cite{brewer1963ratio}.

\subsection*{Why it works}

To understand why the new method works, consider the two stages of
sampling.  The first-stage is the actual network sampling design by
which the sample of people is selected from the hidden population.
The second-stage design selects a resample of people from the network
sample data, using a network sampling design similar to the one used
in the real-world.  The second-stage design, like the first, uses
link-tracing, branches, and is done without-replacement.  The
second-stage design can not be exactly the same as the original design in
every respect.  For example the second-stage design has to use a
smaller sample size than the original, because of the
without-replacement sampling.

The probability that individual $i$ in the
original sample is included in the resample will be called $\phi_i$.  
The ideal is to have the inclusion probability for
a unit at the second stage, given the first stage sample, to be
proportional to it's inclusion probability in the original design.
That is, $\phi_i = c\pi_i$, where $c$ is some constant, which does not
need to be known.

Now let $\hat\mu_{\phi}$ be formula \ref{eq:est} with the exact
resample probabilities $\phi_i$ replacing $f_i$.  If $\phi_i = c\pi_i$,
then $\hat\mu_{\phi} = \hat \mu_{\pi}$, because the constant of
proportionality $c$ is in both the numerator and denominator of
\ref{eq:est} and divides out of the estimator.

  As the number of resamples $T$ gets large the inclusion
frequencies $f_i$ converge in probability to the second-stage
inclusion probabilities $\phi_i$.  This is by the (weak) Law of Large
Numbers for the independent resamples and by the Law of Large Numbers
for Markov chains for the resampling process that traces a few and
removes a few at each step.

It follows that if $\phi_i$ is proportional to $\pi_i$ then $\hat\mu_f$ converges
in probability to $\hat\mu_\phi$.  So if inclusion in the resample
$\phi$ is proportional to inclusion in the original sample $\pi$ then
the estimator we use here $\hat\mu_f$ converges to the general unequal
probability estimator $\hat\mu_{\pi}$.  Since the resampling is fast
computationally, especially with the sampling process approach, we can
readily select a lot of resamples, such as $T = 10,000$ that we use in
the simulations here, and higher values of $T$ like one million are
still fast to compute.  

The approximate part is in how close the second-stage inclusion
probabilities $\phi_i$ are to proportionality with the first-stage
inclusion probabilities $\phi_i$.  This is why it is important that
the resampling design adheres to the main features of the actual
network design, such as network link tracing, branching, and
without-replacement selections.

The use of the second stage sample is different here than in
traditional two-stage sampling or in bootstrap methods.  In each of
those a given second-stage sample is used to make an estimate of a
population value.  If the sampling is with unequal probabilities at
each stage, estimation of the population value from the second-stage sample
requires dividing first by the second-stage inclusion probability $\phi_i$ to
estimate what is in the first stage sample and then then by the first
stage probability $\pi_i$.  Here we use the second stage design solely
to estimate its own inclusion probabilities and we construct the
second-stage design to have those probabilities similar to the
first-stage probabilities.

\subsection*{Background}

Early uses of network sampling to find and study hidden populations in the
typically used snowball sampling types of
methods.  Sample means and proportions were used to summarize data and
implicitly to infer hidden population characteristics.  Reviews of the
early literature can be found in \cite{spreen1992rare},
\cite{heckathorn1997respondent},  and \cite{thompson2002adaptive}.
Unbiased estimates of population values from relatively simple network
sampling designs were obtained by \cite{birnbaum1965design} and
\cite{frank1977survey}.  \cite{frank1994estimating} obtained design
and model based estimates of the size of a hidden population of drug
users with a one-wave snowball sample.  

Starting with \cite{heckathorn1997respondent}, the methodology of
Respondent-Driven-Sampling using dual-incentive coupons was
introduced.  Estimators for these designs based on random walk theory
and assumptions of Markov transitions in the sampling between values
of attribute variables of respondents were given in
\cite{salganik2004sampling}, \cite{heckathorn20076}, and
\cite{volz2008probability}. If a random walk with replacement is run
in a network consisting of a single connected component, the long term
frequency of inclusion of node $i$ is proportional to $d_i$. The
simplest of these estimators divides observed value for an individual
$i$ by degree $d_i$, the number of partners that person reports
having.  In relation to Equation \ref{eq:est} it could be denoted
$\hat\mu_d$.  It has the form of the general unequal probability
estimator with $\pi_i$ approximated by $d_i$.  It is commonly called the
Volz-Heckathorn estimator (VH) or the Salganik-Volz-Heckathorn
estimator.  Following \cite{goel2010assessing}, in this paper we refer
to the VH also as the Current method, because it is the most commonly
used in current practice and its assumptions are also made in the
other commonly used variants.  The estimator of
\cite{salganik2004sampling} uses the VH estimator to estimate degree
and for attribute variables adjusts that with proportions of sample
recruitment links between and within the group with the attribute
and the group without it.  The adjustment is based on the additional
assumption of Markov transitions between attribute states during the
sampling.

The Successive Sampling (SS) estimator of Gile \cite{gile2011improved}
adjusts the VH estimator for samples in which sample size $n$ is a
substantial fraction of population size $N$.  The improvement is based
on the fact that the real sampling is done without replacement.  As
part of the procedure resampling with probability proportional to size
among all units not already in the resample is used to estimate
inclusion probability.  In this way the use of resampling in
\cite{gile2011improved}, although it does not include link-tracing or
branching, is similar in purpose to the use of the resampling in this
paper.  The measure of size of a unit in the successive sampling
method is degree.  When population size is large compared to sample
size the SS esstimator defaults to the VH estimator.

Fellows \cite{fellows2018respondent} introduces the Homophily
Configuration Graph Estimator (HCG), which uses the attribute variable
group membership in a manner similar to SH and adjusts for large
sampling fraction with the SS method.  When population size is large
relative to sample size the HCG estimator defaults to the SH estimator
for estimating the proportion with an attribute and to the VH for
estimating mean degree.  

Confidence interval methods commonly used with RDS designs include the
Salganik bootstrap \cite{salganik2006variance} for SH and VH, the Gile
SS bootstrap \cite{gile2011improved} for SS.  A recent evaluation of
these methods, for means of binary variables and using simulations
based on a statistical network model fitted to RDS data is
\cite{spiller2017evaluating}.  

A different bootstrap approach for RDS data called a Tree Bootstrap is
described in \cite{baraff2016estimating} for making a confidence
interval around the VH estimator.  The bootstrap resamples are
selected by first taking a random sample with replacement of the
sample seeds.  From each of those their sample tree is resampled,
starting with the original seed and tracing links with replacement
except not moving backwards to resample the recruiter at any step.
With each resample, the VH estimator is calculated and the confidence
interval is based on the quantiles of the distribution of the resample
estimates. In making the bootstrap estimates of the population values,
the necessity of dividing by second-stage (resample) inclusion probabilities is
avoided by the equal-probabililty resampling of seeds.  The Tree Bootstrap
confidence intervals have good coverage probabilities but are wide
compared to other methods.  The need for wide intervals in order to
have good coverage probability is not inherent to the Tree Bootstrap
approach but is needed to overcome the bias offet of the VH
estimator.

Estimation methods based on the VH estimator are based on node
degrees.  The SH estimator goes one step further into the sample
network by using the proportion of links within a group to the number
going out from that group.  In contrast the methods in \cite{thompson2006aws},
\cite{crawford2016graphical}, and \cite{crawford2018hidden} use the
full sample network.  Minimally, the sample network includes only
those links within the sample that were used in recruitment, plus the
counterpart edge in the other direction to make the link symmetric.  The
more full version of the sample network uses the set of all links that
connect sample nodes.

RDS methods were evaluated in \cite {goel2010assessing} using a
collection of real populations, of which Project 90 population is the
most relevant to our interests here since it is an actual at-risk,
hidden population in which link-tracing provides the only means of
access and is not based in an institutional setting.  They used
simulations in which the RDS sampling was modeled as with-replacement,
noting that in the field they were normally done without-replacement.
They found the VH and SH estimators performed similarly and focused on
study of the VH estimator, termed the Current Estimator.  For the 13
attribute variables in the study their simulations found actual
confidence interval coverages from 42\% to 65\% for nominally 95\%
confidence intervals.  Design effects were found to be high with the
RDS methods, reflecting the high mean squared errors.

RDS with the VH and SH estimators is examined in \cite{gile20107}
using simulation of the design with and without replacement and with
branching in a variety of model-generated network populations.  They
found the VH estimator generally out-performed the SH estimator and
that without-replacement designs generally resulted in lower variance
and lower bias than with-replacement designs.  They found biases in
the estimates in many conditions.  Between selection of seeds at
random and with probability proportional to degree they found little
difference in the resulting estimators, but stronger seed selection
biases could affect the resulting properties.   

An RDS design with supplement social network stucy, making the overall
design RDS+ in the terminology of this paper, is used in
\cite{young2018network} to study a rural opioid user network in
relation to HIV risk.

In this report the design-adherent estimators are evaluated with
simulations using the reference Project 90 network data on a hidden
population at risk for HIV \cite{potterat1999network}.  Some results
of the present report were described in the earlier ArXiv 
working paper \cite{thompson2018simple}.

\section{Methods}

\subsection*{Confidence intervals} A simple variance estimator to go with $\hat\mu_f$ is

\begin{equation}
\widehat{\textrm var}(\hat\mu) =  \frac{1}{n(n-1)}  \sum_{s}
\left(\frac{ny_i/f_i}{\sum_s (1/f_i)} - \hat\mu_f\right)^2
\end{equation}

The variance estimator is based on a simplified  estimator for
the variance of a Horvitz-Thompson estimator evaluated with 
simulations in \cite{brewer1983sampling}, which has been modified here
 to apply  to the
generalized unequal probability estimator form of $\hat\mu_f$.  
An approximate $1 - \alpha$ confidence interval is then calculated as
$ \hat\mu \pm z \sqrt{\widehat{\textrm var}(\hat\mu)}$, with $z$ the
$1-\alpha/2$ quantile from the standard Normal distribution.  A
different variance estimator was used for these estimators in the
working paper \cite{thompson2018simple}.  In the simulations here the
estimator above gives a slightly higher average confidence interval coverage
probability.

\subsection*{Simulations}

The simulations were done using the entire network data set of 5492
people and 21,644 links from the Colorado Springs Project 90 study on
the heterosexual spread of HIV \cite{potterat1999network}.  The same
data set was used in simulations in \cite {goel2010assessing},
\cite{baraff2016estimating}, and \cite{fellows2018respondent}.  The
data are available to researchers
(https://opr.princeton.edu/archive/p90/).  The links combine drug,
sexual, and social relationships.

For each of the four designs, 1000 samples of target size $n = 1200$
were selected from the 5492 study population.  In RDS and RDS+, 3
coupons were given to each respondent (fewer if the respondent had
fewer than 3 partners).  In SB and SB+, the coupon maximum was 15.
Coupons had an expiration date 28 days from issue.  Seeds (240 or 20\%
of $n$) were selected at random.  The resampling process, like
the original design, used link-tracing, branching, and
without-replacement sampling.   No  coupons were used in the resampling process,
so that the same resampling design was used for each of the four
original designs.  As can
be seen from the RDS sample in Figure \ref{fig:sample} where each
respondent was given no more than 3 coupons with which to, a
without-replacement resample can at no point branch more than 3 in any
case, or up to 4 branches from a re-seed.  For each of the 1000 samples, the design-adherent estimator was
calculated by selecting $T=10,000$ resamples each of target size 400
and averaging the inclusion indicators for each of the 1200 sample
people, giving the frequencies $f_i$ to calculate the estimate
$\hat\mu_f$.

\subsection*{Resampling process versus independent resamples}

Selecting many samples from the sample network data to estimate the
inclusion probabilities can be done with either independent resamples
from seeds to target sample size or using the Markov chain resampling
process.   The resampling processes method is much faster
computationally and is the approach that was used in the simulations.

\begin{figure}
\centering
\includegraphics[width=0.9\linewidth]{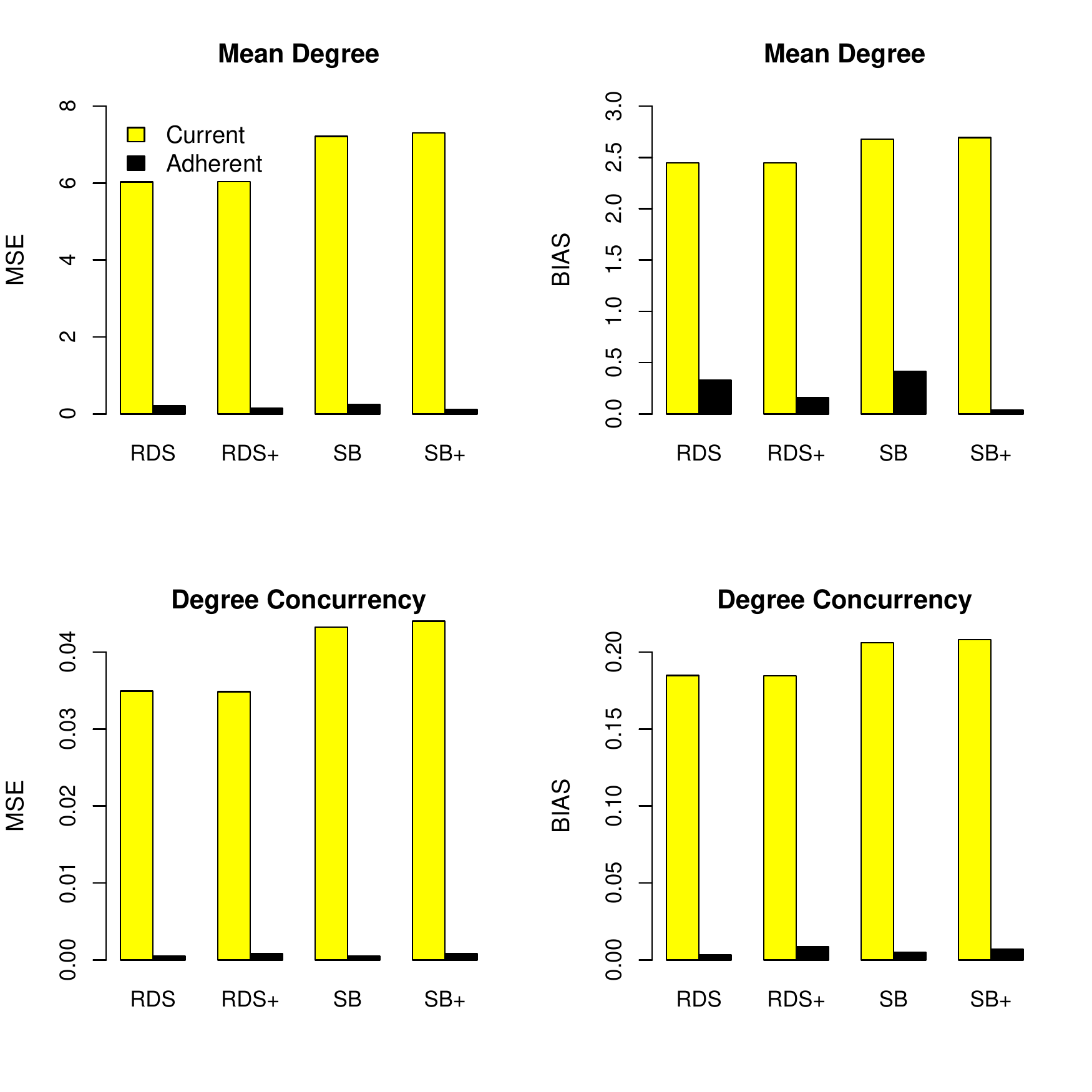}
\caption{The mean square error (left) and bias (right) of the adherent
  estimator (black) is lower than that of the current estimator
  (yellow) for estimating mean degree (top) and degree
  concurrency---proportion of people with two or more partners
  (bottom) for each of four network sampling designs.  The design RDS
  restricts branching to limiting coupons to a maximum of 3.  The
  snowball (SB) design gives respondents as many coupons as their
  number of partners, up to a maximum of 15, allowing almost unlimited
  branching as the sample is selected.  The ``+'' version of each
  design obtains additional social network information from the given
  set of sample members.  }
\label{fig:bar}
\end{figure}

\begin{figure}
\centering
\includegraphics[width=1.0\linewidth]{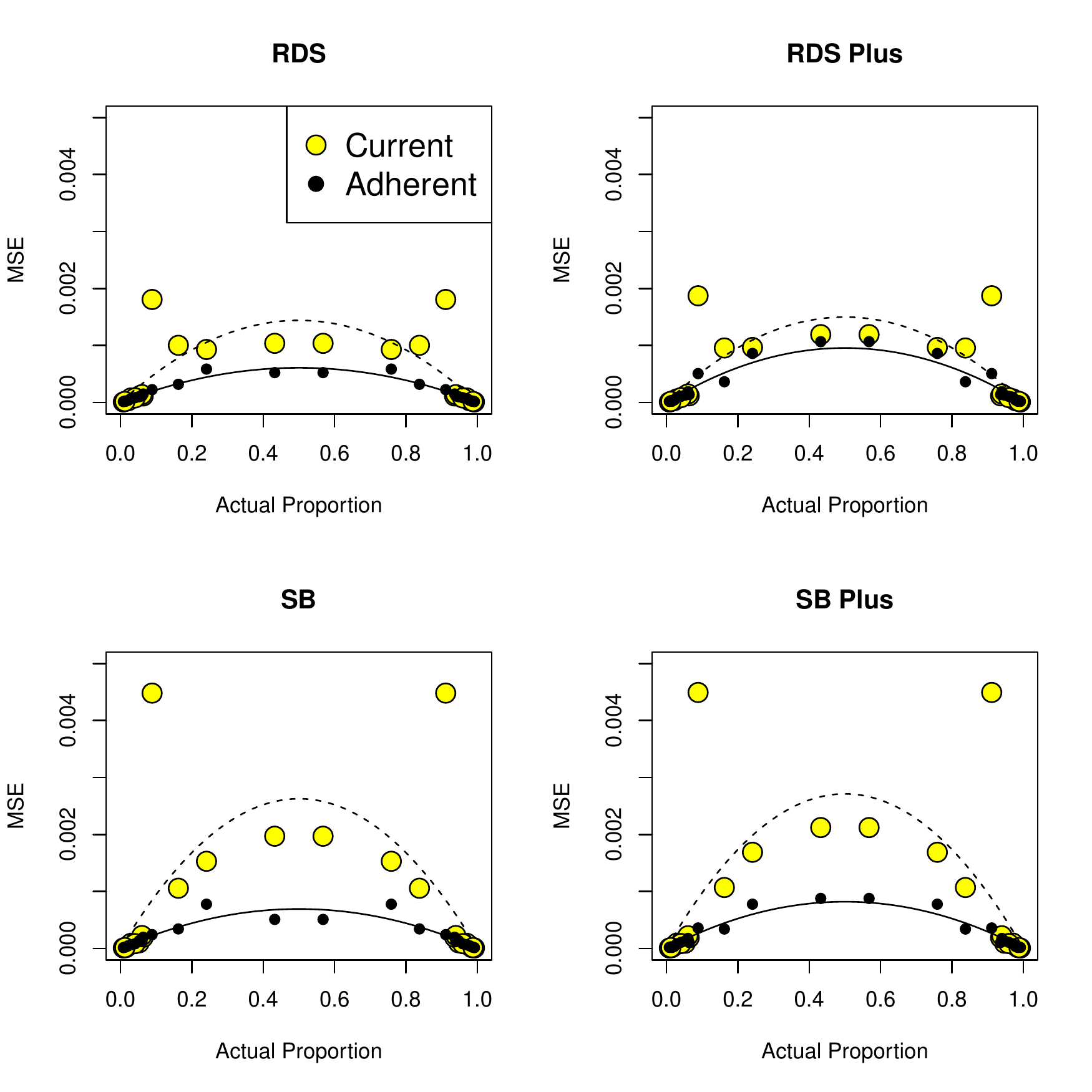}
\caption{Mean squared error of the estimate of population proportion
  for each of 13 individual attributes. The adherent estimator (black)
  is compared to the current estimator (yellow).  To help see the
  pattern, the estimate of the compliment of each attribute is shown.
  The compliment of sex work client, for example, is not-client.  A
  parabolic curve is fitted by weighted least squares to the MSEs of
  the adherent estimator (solid line) and the current estimator
  (dashed line).  The adherent estimator MSEs have a lower fitted
  curve with a better fit with each design.  }
\label{fig:parabola}
\end{figure}

\begin{table}[ht]
\centering
\begin{tabular}{llrcrc}
  \hline
 & & RDS & RDS Plus & SB & SB Plus \\ 
  \hline
 1 & degree & 0.77 & 0.86 & 0.72 & 0.92 \\ 
 2 & nonwhite & 0.95 & 0.86 & 0.92 & 0.88 \\ 
 3 & female & 0.98 & 0.99 & 0.99 & 0.99 \\ 
 4 & sex worker & 0.95 & 0.92 & 0.95 & 0.94 \\ 
 5 & procuring agent & 0.90 & 0.69 & 0.95 & 0.85 \\ 
 6 & sex-worker client & 0.95 & 0.72 & 0.96 & 0.79 \\ 
 7 & dealer & 0.95 & 0.78 & 0.94 & 0.85 \\ 
 8 & cook & 0.78 & 0.65 & 0.79 & 0.67 \\ 
 9 & thief & 0.94 & 0.65 & 0.93 & 0.71 \\ 
 10 & retired & 0.93 & 0.84 & 0.93 & 0.84 \\ 
 11 & homemaker & 0.94 & 0.93 & 0.93 & 0.91 \\ 
 12 & disabled & 0.94 & 0.89 & 0.94 & 0.89 \\ 
 13 & unemployed & 0.95 & 0.94 & 0.95 & 0.94 \\ 
 14 & homeless & 0.88 & 0.67 & 0.86 & 0.67 \\ 
 15 & degree concurrency & 1.00 & 1.00 & 1.00 & 1.00 \\ 
   \hline
\end{tabular}
\caption{Confidence Interval Coverage Probabilities } 
\label{coverageall}
\end{table}

\section{Results}

Two link-related quantities of interest in studies of key populations at risk
for HIV and  other infectious diseases are mean degree and degree
concurrency.  Mean degree is the average number of partners per person
in the population.  Degree concurrency is the proportion of people in
the population who have two or more partners.  A high number for
either of these is an indication that an epidemic could spread rapidly
in the population once it starts there.

The top two plots in Figure \ref{fig:bar} are on estimating mean
degree.   In each plot the four designs, from left
to right, are standard Respondent Driven Sampling (RDS), enhanced
Respondent Driven Sampling (RDS+), Snowball Sampling (SB), and
enhanced Snowball Sampling (SB+).  Respondent Driven Sampling limits
each recruiter to 3 coupons, which which they can recruit up to 3 of
their partners into the study.  Snowball Sampling here has a coupon
limit of 15.

In Figure \ref{fig:bar} the top left plot shows the mean
squared error for the current estimator (yellow bar).  The black bar
next to it is the mean square error for the design-adherent estimator.
Starting with the left-most pair of bars, the mean square error for
the current method (VH estimator) for RDS sampling is
6.03.  The mean square error for the design-adherent estimator, in
which the estimation takes into account the branching and without
replacement in the sampling, the mean square error is 0.21.   The reduction
in mean squared error comes largely from eliminating most of the bias,
as shown in the top right plot of Figure \ref{fig:bar}.  The
actual mean degree in the Project 90 study population is 7.89 partners
per person.  The current estimator underestimates this on average by
2.45 partners (yellow bar).  The adherent estimator overestimates by
0.32 partners on average (black bar).

To understand better what is happening in the estimation we will look
more closely at this one case before going on to the overall pattern
in all the results.  Mean squared error of an estimator is related to
its bias and standard deviation by MSE = (Bias)$^2$ + (SD)$^2$.  Here
the MSE.  Here the squared bias 5.99 accounts for almost all of the
MSE 6.03.  The adherent estimator reduces the squared bias to 0.11
which is about half of its MSE.  So the efficiency gain is due to the
elimination of most of the bias.  The reduction in bias is obtained by
the more accurate estimates of inclusion probabilities using a the
resampling design that adheres to the the branching  and
without-replacement features of the actual survey design, and using
the sample network recruitment data instead of  assumptions
about the hidden population network.  

For estimating mean degree, with the RDS design the relative
efficiency of the design-adherent method,
(MSE($\hat\mu$)/MSE($\hat\mu_d$), is 29.  So with the same design,
adhering to design features in estimation reduces the mean squared
error to 1/29 that of the current method.  With the other designs, the
efficiency gains of the adherent estimation are 41 for RDS Plus, 29
for SB, and 68 for SB plus.  In each case this is accomplished through
the reduction in bias with the design-adherent method.

For estimating the proportion of people in the population with two or
more partners (degree concurrency), the relative efficiencies of the
adherent estimators compared to the current estimators are 72 for RDS,
44 for RDS Plus, 92 for SB, and 55 for SB Plus.  Once again the
reduction in MSE is explained by the reduction in bias.

The Project 90 node data includes 13 individual attribute variables such as sex
worker, client of sex worker, or unemployed.  These are variables 2
through 14 in Table 1.  For an individual, the value is 1 if the
individual has the attribute and 0 otherwise.  The object for
inference for each attribute is to estimate the proportion of people
in the population having that attribute.  For the simulations, missing
values were arbitrarily set to zero so that sample sizes would be the
same for all variables.  

 To help see the pattern in the mean squared errors for estimating the
 population proportions of the 13 attribute variables, the
MSE for each variable is plotted in Figure \ref{fig:parabola} against the actual proportion of
people having that attribute in the Project 90 study population.  For each of the 13 variables, we can
also estimate the proportion for its complement.  The compliment of
``client'', for example, is ``not client''.  The proportion for the
compliment is 1 minus the proportion with the attribute, and the MSE
for estimation the compliment is the same as the the MSE for
estimation of the original variable.  This gives us 28 variables for
each plot in Figure \ref{fig:parabola}, with actual proportions
ranging from near 0 to near 1.  The original variables are on the
left, since the actual proportions are all less than one-half.  The compliment variables provide redundant information but clarify
the pattern in the MSEs.  For each design, the mean squared
errors of the design-adherent estimators fall close to a parabola
shape as a function of the actual proportion.  The parabolas in the
plots have form $MSE = ap(1-p)$ where $p$ is the actual proportion and
$a$ is estimated for each design separately with weighted least
squares, using only the original data.

The MSE with the adherent estimator (black in plots) is lower than that 
of the current estimator in all cases except for some of the ones with
actual proportion near to zero for which the MSE is very small with
either estimator.  While the MSEs of the adherent estimator fall
rather close to the fitted parabola (solid line), the MSEs of the
current estimator are more erratic and the fitted parabola (solid
line) is higher.  The overall higher MSEs and erratic pattern with the
current estimator result from the discrepancies between actual
inclusion probabilities and those used in estimation.     

Confidence interval coverage probabilities for each variable for each
of the 15 variables are given in Table 1.  While the nominal coverage
is 95 percent, the actual coverage probability is assessed in the
simulation as the proportion of the 1000 runs, corresponding to 1000
original samples of size 1200 each, for which the sample confidence
interval contained the true value for the population.  For the values
rounded to 1.0 in the last row of the table, the exact coverage
proportions were 0.999, 0.997, 0.998, and 0.996.  
For estimating mean degree, confidence interval coverage probability
is higher with the 'plus' version than with the regular version of
each design.  For most of the variables, however, the confidence
interval coverage is higher with the regular version of the design.


\section{Discussion and conclusions}

The new estimators obtain better estimates of key population
characteristics from network survey data because the method 
adheres to main features of the actual design, such as branching and
sampling without replacement.  Currently-used estimators are based
instead on the inclusion probabilities of a theoretical design that
does not adhere to the features of the actual survey design and suffer
biases and large mean square errors as a result.  

The advantage of the design-adherent estimators is especially great
for estimating link-related quantities such as mean degree and degree
concurrency, for which the values are strongly related to survey
inclusion probabilities.  Importantly, the link-related variables are
directly related to the network risk of HIV spread in the population.
The elimination of most of the bias through the adherent estimation
method enables confidence intervals of modest width to have coverage
probabilities close to the desired nominal value.

The new estimators work for coupon-restricted
Respondent-Driven-Sampling designs and for freely branching Snowball
Sampling designs.  Current estimators are intended for RDS designs and
not recommended for Snowball designs.  The designs obtaining
information on within-sample links beyond those used in recruitment
(``plus'' designs) did not perform consistently better than their
regular counterparts for estimation, though they may have other uses
such as estimating differential recruitment tendencies.

The results of this study indicate that much added value can be gained
from existing network survey data by re-analysis with the
design-adherent methods, and that for new surveys the new estimation
methods can be recommended.

\section*{Acknowledgements}
This research was supported by Natural Science and Engineering
Research Council of Canada (NSERC) Discovery grant RGPIN327306.  I would like
to thank John Potterat and Steve Muth for making the Project 90 data
available. I would like to express appreciation for the participants
in that study who shared their personal information with the
researchers so that it could be made available in anonymized form to
the research community and contribute to a solution to HIV and
addiction epidemics and to basic understanding of social networks.




\section*{APPENDIX}

The appendix includes (1) additional description of the two approaches, repeated
samples and sampling process to calculating the inclusion frequncies
$f_i$ which estimate the real-world inclusion probabilities $p_i$; (2)
additional details and variations on the estimators and variance
estimators; (3)  supplemental tables that include the numbers behind
the figures and additional values such as expected values of variance
estimators and mean confidence interval half-widths.

\subsection*{Repeated samples and sampling process}

The inclusion frequencies $f_i$ are calculated by re-sampling the
sample network many times using a design similar to the original
design used to select the members of the hidden population from the
real world.  Two approaches to the resampling are to repeatedly select
independent resamples, each from seeds to target sample size, and to
select the sequences of resamples using a Markov-chain resampling
process.  


In the simulations of the paper only the sampling process approach was
used.  The repeated-sample approach is illustrated here first as
understanding that makes the sampling-process approach easier to
understand.

Figure 1 of the text shows an RDS sample of 1200 people selected from
the Colorado Springs network population of sex workers, drug users,
and partners of each (\cite{potterat1999network}).  
The target sample size of the resamples is 400.

The simulation study selected 1000 samples of 1200 from the
population, for each of the four designs RDS, RDS+, SB, and SB+.  The
simulation study used the sampling process method, which is
computationally very much faster than the independently repeated
samples.

Given the network sample obtained from the real world
network sampling design, we obtain a sequence of re-samples  
\[
\{S_1, S_2, S_3, ...,S_T \}
\] 
from the network data using a fast-sampling process 
similar to the original design.  $T$ is the number of iterations.

For unit $i \in U_s$, there is a sequence of indicator random variables:
\[
\{Z_{i1}, Z_{i2}, Z_{i3},..., Z_{iT} \}
\]
where $ Z_{it} = 1$ if $i\in S_t $ and  $Z_{it} = 0$ if $ i \notin
S_t$, for $t = 1, 2, ..., T$, the number of iterations of the sampling
process.  

The average 
\[
f_i = \frac{1}{T} \sum_{i = 1}^T Z_i
\]
is used as an estimate of the relative inclusion probability  of unit
$i$ in the similar design used to obtain the data from the real
world.  If the real-world network design is done without replacement,
then the fast-sampling process is also carried out without
replacement.  

In the repeated-sample approach, each sample in the sequence proceeds from
selection of seeds to target sample size.  With this approach the
samples in the sequence $\{S_1, S_2, S_3, ...,S_T \}$ are independent
of each other.  

In the sampling-process approach, each sample $S_t$ is selected
dependent on the one before it, $S_{t-1}$.  To get from $S_t$ to
$S_{t-1}$ we probabilistically trace links out from $S_t$, randomly
drop some nodes from $S_t$, and may with low probability select one or
more new seeds.  Advantages of the sampling-process approach are
first, that the computation can be made very fast.  Second, the
sampling process is fast-mixing and once it reaches it's stationary
distribution every subsequent sample $S_t$ is in that distribution.
The stationary distribution of the sequence of samples represents a
balance between the re-seeding distribution, which can be kept small
with a low rate of re-seeding, and the design tendencies arising from
the link-tracing and the without-replacement nature of the selections.

The sampling process is without-replacement in that a node in the
sample can not be selected again while it is in the sample.  Once it
has been removed from the sample it can be selected anew at any time.  With
the sampling-process design, the sequence of fast samples $S_1, S_2, ...$
forms a Markov chain of sets, with the probability of set $S_t$
depending only on the previous set $S_{t-1}$.  



In the simulations the repeated-sample design selects initial seeds using
Bernoulli sampling.  A low rate of re-seeding is used, mainly to ensure that
the growth of the sample can not get stuck before target sample size
is reached.  A medium rate of link-tracing is used.  Links out
are traced with independent Bernoulli selections.  Because no coupons
were used in the re-sampling, each selected node can continue to
recruit without time limit.

The sampling process can use a high tracing rate because
removals offset the tracing to maintain a stochastic balance around
target sample size.
A relatively high rate of ongoing reseeding rate is used so that the
process does not get locked out of any components.  Whenever the sample goes
above target sample size, the sample is randomly thinned with
removals, with probability of removal set to make expected sample size
back within target at the next step.  All these features make the
process very fast mixing.

Specifically, in the examples we trace the links out from the current
sample $S_t$ independently, each with probability $p$.  Nodes are
removed from the sample independently with probability $q$.  The
removal probability $q$ is set adaptively to be $q_t = (n_t - n_{\rm
  target})/n_t$ if $n_t > n_{\rm target}$ and $q_t = 0$ otherwise, so
that sample size fluctuates around its target during iterations.
Sampling is without replacement in that a node in $S_t$ is not
reselected while it remains in fast sample, but it may be reselected
at any time after it is removed from the fast sample.  The re-seeding
rate can be low because the seeds at the beginning of each sample
serve to get the sample into enough components.

In the repeated-samples design, the seeding rate $p_s =
0.0167$; the tracing rate is $p = 0.05$, and the re-seeding rate is
$p_r = 0.001$.  The sampling-process design  uses no
initial seeds, relying on the re-seeding to initialize the process and
bring it quickly into its stationary distribution.  The removal rate
$q$ is set adaptively as described above to keep the process in
fluctuation around its target sample size.  The re-seeding rate is $p_r
= 0.01$.  Even though the re-seeding rate is relatively high, at each
step most added nodes are added by tracing links, because that rate
is so much higher.  The re-seeding serves to keep the process from
getting permanently locked out of any component through removals.

If the real-world survey sampling is done with replacement, one can
use a re-sampling  design that is with replacement.  An
advantage of this is that a target sample size for the fast design can
be used that equals the actual sample size used in obtaining the
data.  However, in most cases the real design is done without
replacement.

Sampling processes of these types are discussed in 
\cite{thompson2017adaptive} and \cite{thompson2015fast} for their
potential uses as measures of network exposure of a
node, or a measure of network centrality, or a predictive indicator of
regions of a network where an epidemic might next explode.
Calculation of the statistic $f_i$ for each unit in the network sample
can be used as an index of the network exposure of that unit.  A high value
of $f_i$ indicates the unit has high likelihood of being reached by a
network sample such as ours.  It will also have a relatively high
likelihood of being reached by a virus, such as HIV, that spreads on
the same type of links by a link-tracing process that is broadly
similar.  A given risk behavior will be more risky for a person with
high network exposure.  For a person in a less well connected part of
the network, the same behavior carries lower risk.  Since a purpose of
the surveys is to identify risk characteristics, an index of network
exposure measures another dimension of that risk, beyond the individual
behavior and health measures.  Here, however, are interested in their
usefulness for estimating population characteristics based on
link-tracing network sampling designs.

\subsection*{Estimators}

This supplemental section contains additional detail on estimation
formulas.  It includes estimation when the original design is carried
out with replacement and estimation of a ratio.  

Network sampling designs select units with unequal
probabilities.  With unequal probability sampling designs, sample
means and sample proportions do not provide unbiased estimates of
their corresponding population means and proportions.  

To estimate the mean of variable $y$ with an unequal-probability
sampling design, the generalized unequal probability estimator has the
form 
\begin{equation}
\hat\mu_{\pi} = \frac{\sum_s (y_i/\pi_i)}{\sum_s (1/\pi_i)}
\end{equation}
where
$\pi_i$ is inclusion probability of unit $i$.

With the network sampling designs of interest here, the inclusion
probabilities $pi_i$ are not known and can not be calculated from the
sample data.  To circumvent this problem the Volz-Heckathorn Estimator
uses degree, or self-reported number of partners,  to approximate inclusion probability:
\begin{equation}
\hat\mu_{d} = \frac{\sum_s (y_i/d_i)}{\sum_s (1/d_i)}
\end{equation}
in which $d_i$ is the degree, the number of self-reported partners, of person $i$.

The rationale for this approximation is that if the sampling design is
a random walk with replacement, or several independent random walks with replacement and the population
network is connected, then the selection probabilities of the random
walk design will converge over time to be proportional to the $d_i$.
Here connected means that each node in the population can be reached
from any other node by some path, or chain of links, so that the
population network consists of only one connected component.  Biases
in this estimator result from the use of without-replacement sampling
in the real-world designs, the use of coupon numbers $k$ greater than
1 making the design different from a random walk, population networks
being not connected into a single component, or slow mixing due to
specifics of the
population network structure.    

The  design-adherent estimator, with a non-replacement
sampling design, is 
\begin{equation}
\hat\mu_f = \frac{\sum_{i \in s} (y_i/f_i)}{\sum_{i \in s} (1/f_i)}
\end{equation}
where
$f_i$ is the inclusion frequency of unit $i$ in the resampling process run
on the sample network data.

A simple variance estimator to go with the adherent estimator is
\begin{equation}
\widehat{\textrm var}(\hat\mu_f) = \frac{1}{(\sum_{i \in s}
  1/f_i)^2}\sum_{i \in s} \frac{(y_i - \hat\mu_f)^2}{f_i^2}
\label{eq:hatvar1}
\end{equation}

Another simple variance estimator is 
\begin{equation}
\widehat{\textrm var}(\hat\mu) =  \frac{1}{n(n-1)}  \sum_{s}
\left(\frac{ny_i/f_i}{\sum_s (1/f_i)} - \hat\mu_f\right)^2
\label{eq:hatvar2}
\end{equation}

\medskip
\medskip

An approximate $1 - \alpha$ confidence interval is then calculated as 
\begin{equation}
\hat\mu_f \pm z \sqrt{\widehat{\textrm var}(\hat\mu_f)}
\end{equation}
with $z$ the $1-\alpha/2$ quantile from the standard Normal distribution.

The variance estimator \ref{eq:hatvar1} is based on, and simplified from, the Taylor
series linear approximation theory for generalized unequal probability
estimator. 
Linearization leads to the estimator of the variance of the
generalized estimator 
\begin{equation}
\widehat{\textrm var}(\hat\mu_{\pi}) = \frac{1}{(\sum_{i \in s}
  1/\pi_i)^2}\sum_{i \in s} \sum_{j \in s} \check\Delta_{ij}\frac{(y_i
  - \hat\mu_{\pi})}{\pi_i}\frac{(y_j - \hat\mu_{\pi})}{\pi_j}
\end{equation}
where 
\[
\check\Delta_{ij} = \frac{\pi_{ij} - \pi_i\pi_j}{\pi_{ij}}
\]
where 
$\pi_{ij}$ is the joint inclusion probability for units $i$ and $j$.  
A good discussion of the approach is found in \cite{sarndal2003model},
with this variance estimator on p. 178 of that work.

\medskip

The variance estimator \ref{eq:hatvar2} is based on the idea that if
the sum $\hat\mu_f = \sum_{i \in s} (y_i/f_i)/\sum (1/f_i)$ estimates
$\mu$ then each piece $(y_i/f_i)/\sum (1/f_i)$ estimates $\mu/n$ and
so $t_i = n(y_i/f_i)/\sum (1/f_i)$ would be an estimate of $\mu$, for
$i=1, ..., n$.  Ignoring the dependence from the without-replacement
sampling and treating $t_1, ..., t_n$ as uncorrelated, then
$\hat\mu_f$ is the sample mean of the $t_i$ and \ref{eq:hatvar2} is
their sample variance divided by sample size.  

In simulations both \ref{eq:hatvar1} and \ref{eq:hatvar2} give decent
variance estimates and confidence intervals.  The coverage probability
tended to be modestly better with \ref{eq:hatvar2}, and that is the
one used in the simulations of the report.  

Consider an estimator of the variance using the full variance
expression with  the fast-sample
frequencies $f_i$ in place of the $\pi_i$ and, in place of the joint
inclusion probability $\pi_{ij}$, the frequency $f_{ij}$ of inclusion
of inclusion of both units $i$ and $j$ in the fast sampling process.
This would give 
\begin{equation}
\widehat{\textrm var}(\hat\mu_f) = \frac{1}{(\sum_{i \in s}
  1/f_i)^2}\sum_{i \in s} \sum_{j \in s} \hat\Delta_{ij}\frac{(y_i - \hat\mu_f)}{f_i}\frac{(y_j - \hat\mu_f)}{f_j}
\end{equation}
where 
\[
\hat\Delta_{ij} = \frac{f_{ij} - f_if_j}{f_{ij}}
\]

\medskip

The double sum in the variance estimate expression will have
$n(n-1)/2$ terms in which $i \ne j$.  The most influential of these
terms are the ones in which the joint frequency of inclusion $f_{ij}$
is relatively large.  Because of the link tracing in the fast sampling
process, sample unit pairs with a direct link between them will tend
occur together more frequently than those without a direct link.  An
estimator using only those pairs with known links between them in the
sample data would be 

\begin{equation}
\widehat{\textrm var}(\hat\mu_f) = \frac{1}{(\sum_{i \in s}
  1/f_i)^2} \left(\sum_{i \in U_s} (f_i - 1)\frac{(y_i -
    \hat\mu_f)^2}{f_i}  +  \sum_{(i,j) \in E_s} \hat\Delta_{ij}\frac{(y_i - \hat\mu_f)}{f_i}\frac{(y_j - \hat\mu)}{f_j}\right)
\end{equation}
where $E_s$ is the sample edge set.  That is, $E_s$ consists of the
known edges $(i,j)$ between pairs of units in the sample data.  In
general the size of the sample edge set $E_s$ will be much smaller
that the $n^2$ possible sample node pairings $(i, j)$, or the
$n(n-1)/2$ pairings with $i\ne j$, where $n$ is the sample size.  

A further simplification and approximation for estimating the variance
of the estimator is to use only the diagonal terms, that is, 
\begin{equation}
\widehat{\textrm var}(\hat\mu_f) = \frac{1}{(\sum_{i \in s}
  1/f_i)^2} \sum_{i \in s} (1 - f_i)\frac{(y_i -
    \hat\mu_f)^2}{f_i} 
\end{equation}

Dropping the coefficients $(1-f_i)$, each of which is less than or
equal to one, gives an estimate of variance that is larger, leading to
wider, more conservative confidence intervals.

If the real-world network sampling design and correspondingly the 
re-sampling process are  with-replacement, the estimator of $\mu$ is  
\begin{equation}
\hat\mu_f =  \frac{\sum_{i \in s}
    (m_{i}y_i/g_i)}{\sum_{i \in s}
(m_{i}/g_i)}
\end{equation}
in which $m_i$ is the number of times unit i is selected in the real
design and $g_i$ is the average number of selection counts of unit
$i$ in the fast sampling process.

If the real-world sampling design is with replacement, the re-sampling
process can be done with replacement.  In that case  let $m_t(i)$ be the number of times
node $i$ is selected at iteration $t$.  The quantity $g_i = (1/t)
\sum_{s=1}^t m_t(i)$, the average number of selections up to
iteration $t$, estimates the expected number of selections for node
$i$ under the with-replacement design at any given iteration $t$.

With a with-replacement fast design the corresponding variance
estimator is
\begin{equation}
\widehat{\textrm var}(\hat\mu_f) =   \frac{1}{(\sum_{i \in s}
  m_{i}/g_i)^2}\sum_{i \in s} \frac{m_{i}(y_i - \hat\mu_f)^2}{g_i^2}
\end{equation}

If $x_i$ is another variable, an estimator of the ratio
$R=\mu_y/\mu_x$ of the mean of
$y$ to the mean of $x$ is
\begin{equation}
\hat R = \frac{\sum_{i \in s} y_i/f_i}{\sum_{i\in s} x_i/f_i}
\end{equation}
with simple variance estimator
\begin{equation}
\widehat{\textrm var}(\hat R) = \frac{1}{(\sum_{i \in s}
  x_i/f_i)^2}\sum_{i \in s} \frac{(y_i - x_i\hat\mu_f)^2}{f_i^2}
\end{equation}



\section*{Supplemental Tables}

Tables S1-S4 give the numbers behind the figures in the paper.  In
addition to the 13 attribute variables in the node data file of the
Colorado Springs Project 90 data, the tables include two variables
whose values are calculated from the link data file.  These are
degree, the number of partners a person has, and ``deg2plus'', an
indicator of whether the person has two or more partners.  The
population proportion of people with two or more partners is also
referred to as degree concurrency.  

The column ``actual'' gives the population mean or proportion for each
variable.  ``E.est'' is the mean value of the estimator.  Bias is
E.est - actual.  The standard deviation ``sd'' is $\sqrt{{\rm
var}(\hat\mu)}$ for the given estimator.  The mean squared error
``mse'' is ${\rm E}((\hat\mu -\mu)^2)$.  The relative efficiency
``eff'' is ${\rm E}((\hat\mu_d -\mu)^2)/{\rm E}((\hat\mu_f -\mu)^2)$
for the current estimator $\hat\mu_d$.  For the sample mean $\bar y$
the relative efficiency is similarly defined with the mean squared
error of $\bar y$ in the numerator and that of the adherent estimator
in the denominator.  The relative bias is the ratio of absolute
biases, with the bias of the adherent estimator in the denominator.

Tables S5-S8 give confidence interval coverage probability for nominal
95 percent confidence intervals.  They expand on the information in
the text by giving the average half-width of the interval for each
variable.  Since the intervals are of the symmetric form estimate
$\pm$ half-width of interval, it is natural to look at the average
half-width in relation to the actual value of what is being
estimated.  Coverage probability is the proportion of simulation runs
for which the interval covers the true value.  


\begin{table}[ht]
\centering
\caption{RDS Table} 
\begin{tabular}{lrrrrrrr}
  \hline
Adherent & actual & E.est & bias & sd & mse & eff & rbias \\ 
  \hline
degree & 7.88 & 8.21 & 0.327829 & 0.320172 & 0.209982 & 1.00 & 1.00 \\ 
  nonwhite & 0.24 & 0.25 & 0.010668 & 0.021759 & 0.000587 & 1.00 & 1.00 \\ 
  female & 0.43 & 0.43 & 0.001866 & 0.022826 & 0.000525 & 1.00 & 1.00 \\ 
  worker & 0.05 & 0.06 & 0.003409 & 0.009484 & 0.000102 & 1.00 & 1.00 \\ 
  procurer & 0.02 & 0.02 & 0.001456 & 0.004622 & 0.000023 & 1.00 & 1.00 \\ 
  client & 0.09 & 0.09 & 0.000230 & 0.015043 & 0.000226 & 1.00 & 1.00 \\ 
  dealer & 0.06 & 0.07 & 0.005925 & 0.010702 & 0.000150 & 1.00 & 1.00 \\ 
  cook & 0.01 & 0.01 & -0.000019 & 0.003939 & 0.000016 & 1.00 & 1.00 \\ 
  thief & 0.02 & 0.02 & 0.001522 & 0.006084 & 0.000039 & 1.00 & 1.00 \\ 
  retired & 0.03 & 0.03 & 0.000644 & 0.008213 & 0.000068 & 1.00 & 1.00 \\ 
  homemakr & 0.06 & 0.06 & -0.000079 & 0.010703 & 0.000115 & 1.00 & 1.00 \\ 
  disabled & 0.04 & 0.04 & 0.001477 & 0.009087 & 0.000085 & 1.00 & 1.00 \\ 
  unemploy & 0.16 & 0.17 & 0.006565 & 0.016631 & 0.000320 & 1.00 & 1.00 \\ 
  homeless & 0.01 & 0.01 & 0.000673 & 0.004938 & 0.000025 & 1.00 & 1.00 \\ 
  deg2plus & 0.82 & 0.82 & 0.003064 & 0.021839 & 0.000486 & 1.00 & 1.00 \\ 
    \hline
Current & actual & E.est & bias & sd & mse & eff & rbias \\ 
  \hline
  degree & 7.88 & 5.44 & -2.447003 & 0.215896 & 6.034435 & 28.74 & 7.46 \\ 
  nonwhite & 0.24 & 0.26 & 0.021276 & 0.021866 & 0.000931 & 1.58 & 1.99 \\ 
  female & 0.43 & 0.41 & -0.023358 & 0.022209 & 0.001039 & 1.98 & 12.52 \\ 
  worker & 0.05 & 0.05 & -0.004432 & 0.009785 & 0.000115 & 1.14 & 1.30 \\ 
  procurer & 0.02 & 0.01 & -0.003378 & 0.003509 & 0.000024 & 1.01 & 2.32 \\ 
  client & 0.09 & 0.13 & 0.038526 & 0.017978 & 0.001808 & 7.99 & 167.37 \\ 
  dealer & 0.06 & 0.06 & 0.001224 & 0.010642 & 0.000115 & 0.77 & 0.21 \\ 
  cook & 0.01 & 0.01 & -0.001382 & 0.002870 & 0.000010 & 0.65 & 71.11 \\ 
  thief & 0.02 & 0.02 & -0.000867 & 0.005624 & 0.000032 & 0.82 & 0.57 \\ 
  retired & 0.03 & 0.03 & 0.003194 & 0.008281 & 0.000079 & 1.16 & 4.96 \\ 
  homemakr & 0.06 & 0.05 & -0.008379 & 0.008309 & 0.000139 & 1.22 & 105.95 \\ 
  disabled & 0.04 & 0.04 & -0.004888 & 0.007583 & 0.000081 & 0.96 & 3.31 \\ 
  unemploy & 0.16 & 0.13 & -0.028841 & 0.013125 & 0.001004 & 3.14 & 4.39 \\ 
  homeless & 0.01 & 0.01 & -0.000774 & 0.004193 & 0.000018 & 0.73 & 1.15 \\ 
  deg2plus & 0.82 & 0.64 & -0.184948 & 0.027210 & 0.034946 & 71.86 & 60.35 \\ 
    \hline
$\bar y$ & actual & E.est & bias & sd & mse & eff & rbias \\ 
  \hline
  degree & 7.88 & 14.32 & 6.435291 & 0.235165 & 41.468275 & 197.48 & 19.63 \\ 
  nonwhite & 0.24 & 0.28 & 0.040718 & 0.017822 & 0.001976 & 3.36 & 3.82 \\ 
  female & 0.43 & 0.47 & 0.033170 & 0.011699 & 0.001237 & 2.36 & 17.78 \\ 
  worker & 0.05 & 0.09 & 0.041124 & 0.006060 & 0.001728 & 17.01 & 12.06 \\ 
  procurer & 0.02 & 0.03 & 0.015914 & 0.003251 & 0.000264 & 11.24 & 10.93 \\ 
  client & 0.09 & 0.07 & -0.014696 & 0.007013 & 0.000265 & 1.17 & 63.84 \\ 
  dealer & 0.06 & 0.12 & 0.054420 & 0.006879 & 0.003009 & 20.11 & 9.19 \\ 
  cook & 0.01 & 0.01 & 0.001495 & 0.002406 & 0.000008 & 0.52 & 76.94 \\ 
  thief & 0.02 & 0.04 & 0.014992 & 0.003999 & 0.000241 & 6.12 & 9.85 \\ 
  retired & 0.03 & 0.03 & -0.000374 & 0.004017 & 0.000016 & 0.24 & 0.58 \\ 
  homemakr & 0.06 & 0.07 & 0.007123 & 0.005985 & 0.000087 & 0.76 & 90.07 \\ 
  disabled & 0.04 & 0.06 & 0.014588 & 0.005215 & 0.000240 & 2.83 & 9.87 \\ 
  unemploy & 0.16 & 0.25 & 0.090174 & 0.010090 & 0.008233 & 25.75 & 13.73 \\ 
  homeless & 0.01 & 0.02 & 0.003934 & 0.002662 & 0.000023 & 0.91 & 5.85 \\ 
  deg2plus & 0.82 & 0.93 & 0.110389 & 0.007364 & 0.012240 & 25.17 & 36.02 \\ 
   \hline
\end{tabular}
\label{rdstable}
\end{table}

\begin{table}[ht]
\centering
\caption{RDS Plus Table} 
\begin{tabular}{lrrrrrrr}
  \hline
name & actual & E.est & bias & sd & mse & eff & rbias \\ 
  \hline
degree & 7.88 & 7.73 & -0.155916 & 0.349297 & 0.146318 & 1.00 & 1.00 \\ 
  nonwhite & 0.24 & 0.22 & -0.017929 & 0.023219 & 0.000861 & 1.00 & 1.00 \\ 
  female & 0.43 & 0.44 & 0.012681 & 0.030118 & 0.001068 & 1.00 & 1.00 \\ 
  worker & 0.05 & 0.05 & 0.001842 & 0.011247 & 0.000130 & 1.00 & 1.00 \\ 
  procurer & 0.02 & 0.01 & -0.000864 & 0.004598 & 0.000022 & 1.00 & 1.00 \\ 
  client & 0.09 & 0.07 & -0.016718 & 0.015109 & 0.000508 & 1.00 & 1.00 \\ 
  dealer & 0.06 & 0.06 & -0.006506 & 0.009885 & 0.000140 & 1.00 & 1.00 \\ 
  cook & 0.01 & 0.01 & -0.000576 & 0.004387 & 0.000020 & 1.00 & 1.00 \\ 
  thief & 0.02 & 0.02 & -0.003073 & 0.005948 & 0.000045 & 1.00 & 1.00 \\ 
  retired & 0.03 & 0.03 & -0.002445 & 0.009237 & 0.000091 & 1.00 & 1.00 \\ 
  homemakr & 0.06 & 0.06 & 0.001148 & 0.013752 & 0.000190 & 1.00 & 1.00 \\ 
  disabled & 0.04 & 0.04 & -0.000714 & 0.010393 & 0.000109 & 1.00 & 1.00 \\ 
  unemploy & 0.16 & 0.16 & -0.001906 & 0.019018 & 0.000365 & 1.00 & 1.00 \\ 
  homeless & 0.01 & 0.01 & -0.000525 & 0.005693 & 0.000033 & 1.00 & 1.00 \\ 
  deg2plus & 0.82 & 0.81 & -0.008583 & 0.026654 & 0.000784 & 1.00 & 1.00 \\ 
    \hline
Current & actual & E.est & bias & sd & mse & eff & rbias \\ 
  \hline
  degree & 7.88 & 5.43 & -2.447134 & 0.222663 & 6.038043 & 41.27 & 15.70 \\ 
  nonwhite & 0.24 & 0.26 & 0.022299 & 0.021640 & 0.000966 & 1.12 & 1.24 \\ 
  female & 0.43 & 0.41 & -0.026400 & 0.022246 & 0.001192 & 1.12 & 2.08 \\ 
  worker & 0.05 & 0.05 & -0.005131 & 0.009701 & 0.000120 & 0.93 & 2.78 \\ 
  procurer & 0.02 & 0.01 & -0.003337 & 0.003546 & 0.000024 & 1.08 & 3.86 \\ 
  client & 0.09 & 0.13 & 0.039280 & 0.018162 & 0.001873 & 3.69 & 2.35 \\ 
  dealer & 0.06 & 0.07 & 0.001805 & 0.010718 & 0.000118 & 0.84 & 0.28 \\ 
  cook & 0.01 & 0.01 & -0.001409 & 0.002686 & 0.000009 & 0.47 & 2.45 \\ 
  thief & 0.02 & 0.02 & -0.000646 & 0.005717 & 0.000033 & 0.74 & 0.21 \\ 
  retired & 0.03 & 0.03 & 0.002633 & 0.007536 & 0.000064 & 0.70 & 1.08 \\ 
  homemakr & 0.06 & 0.05 & -0.008574 & 0.008788 & 0.000151 & 0.79 & 7.47 \\ 
  disabled & 0.04 & 0.04 & -0.005409 & 0.007625 & 0.000087 & 0.81 & 7.58 \\ 
  unemploy & 0.16 & 0.13 & -0.027848 & 0.013469 & 0.000957 & 2.62 & 14.61 \\ 
  homeless & 0.01 & 0.01 & -0.000534 & 0.004350 & 0.000019 & 0.59 & 1.02 \\ 
  deg2plus & 0.82 & 0.64 & -0.184722 & 0.027578 & 0.034883 & 44.49 & 21.52 \\ 
    \hline
$\bar y$ & actual & E.est & bias & sd & mse & eff & rbias \\ 
  \hline  
  degree & 7.88 & 14.32 & 6.435271 & 0.243040 & 41.471781 & 283.44 & 41.27 \\ 
  nonwhite & 0.24 & 0.28 & 0.040980 & 0.016936 & 0.001966 & 2.28 & 2.29 \\ 
  female & 0.43 & 0.46 & 0.032076 & 0.011883 & 0.001170 & 1.10 & 2.53 \\ 
  worker & 0.05 & 0.09 & 0.040567 & 0.005974 & 0.001681 & 12.94 & 22.02 \\ 
  procurer & 0.02 & 0.03 & 0.016270 & 0.003400 & 0.000276 & 12.62 & 18.84 \\ 
  client & 0.09 & 0.07 & -0.015001 & 0.006980 & 0.000274 & 0.54 & 0.90 \\ 
  dealer & 0.06 & 0.12 & 0.054818 & 0.007238 & 0.003057 & 21.83 & 8.43 \\ 
  cook & 0.01 & 0.01 & 0.001452 & 0.002248 & 0.000007 & 0.37 & 2.52 \\ 
  thief & 0.02 & 0.04 & 0.014903 & 0.003828 & 0.000237 & 5.28 & 4.85 \\ 
  retired & 0.03 & 0.03 & -0.000851 & 0.003713 & 0.000015 & 0.16 & 0.35 \\ 
  homemakr & 0.06 & 0.07 & 0.007300 & 0.006528 & 0.000096 & 0.50 & 6.36 \\ 
  disabled & 0.04 & 0.06 & 0.014599 & 0.005357 & 0.000242 & 2.23 & 20.45 \\ 
  unemploy & 0.16 & 0.25 & 0.090785 & 0.010585 & 0.008354 & 22.87 & 47.62 \\ 
  homeless & 0.01 & 0.02 & 0.004171 & 0.002799 & 0.000025 & 0.77 & 7.95 \\ 
  deg2plus & 0.82 & 0.93 & 0.110456 & 0.007463 & 0.012256 & 15.63 & 12.87 \\ 
   \hline
\end{tabular}
\label{rdsptable}
\end{table}

\begin{table}[ht]
\centering
\caption{SB Table}
\begin{tabular}{lrrrrrrr}
  \hline
name & actual & E.est & bias & sd & mse & eff & rbias \\ 
  \hline
degree & 7.88 & 8.30 & 0.415435 & 0.274937 & 0.248177 & 1.00 & 1.00 \\ 
  nonwhite & 0.24 & 0.26 & 0.017506 & 0.021710 & 0.000778 & 1.00 & 1.00 \\ 
  female & 0.43 & 0.43 & 0.000860 & 0.022582 & 0.000511 & 1.00 & 1.00 \\ 
  worker & 0.05 & 0.06 & 0.006121 & 0.009902 & 0.000136 & 1.00 & 1.00 \\ 
  procurer & 0.02 & 0.02 & 0.002969 & 0.004545 & 0.000029 & 1.00 & 1.00 \\ 
  client & 0.09 & 0.09 & 0.004574 & 0.014924 & 0.000244 & 1.00 & 1.00 \\ 
  dealer & 0.06 & 0.07 & 0.009554 & 0.010335 & 0.000198 & 1.00 & 1.00 \\ 
  cook & 0.01 & 0.01 & 0.000158 & 0.003987 & 0.000016 & 1.00 & 1.00 \\ 
  thief & 0.02 & 0.02 & 0.002823 & 0.006376 & 0.000049 & 1.00 & 1.00 \\ 
  retired & 0.03 & 0.03 & 0.000947 & 0.007986 & 0.000065 & 1.00 & 1.00 \\ 
  homemakr & 0.06 & 0.06 & -0.001269 & 0.010867 & 0.000120 & 1.00 & 1.00 \\ 
  disabled & 0.04 & 0.04 & 0.001743 & 0.008995 & 0.000084 & 1.00 & 1.00 \\ 
  unemploy & 0.16 & 0.17 & 0.009528 & 0.015928 & 0.000344 & 1.00 & 1.00 \\ 
  homeless & 0.01 & 0.01 & 0.000704 & 0.004722 & 0.000023 & 1.00 & 1.00 \\ 
  deg2plus & 0.82 & 0.83 & 0.005025 & 0.021135 & 0.000472 & 1.00 & 1.00 \\ 
    \hline
Current & actual & E.est & bias & sd & mse & eff & rbias \\ 
  \hline
  degree & 7.88 & 5.20 & -2.679280 & 0.198895 & 7.218099 & 29.08 & 6.45 \\ 
  nonwhite & 0.24 & 0.27 & 0.032360 & 0.022007 & 0.001531 & 1.97 & 1.85 \\ 
  female & 0.43 & 0.39 & -0.038747 & 0.021683 & 0.001971 & 3.86 & 45.07 \\ 
  worker & 0.05 & 0.05 & -0.001975 & 0.009839 & 0.000101 & 0.74 & 0.32 \\ 
  procurer & 0.02 & 0.01 & -0.002209 & 0.003401 & 0.000016 & 0.56 & 0.74 \\ 
  client & 0.09 & 0.15 & 0.063940 & 0.019839 & 0.004482 & 18.40 & 13.98 \\ 
  dealer & 0.06 & 0.07 & 0.008865 & 0.010872 & 0.000197 & 0.99 & 0.93 \\ 
  cook & 0.01 & 0.01 & -0.001520 & 0.002682 & 0.000010 & 0.60 & 9.62 \\ 
  thief & 0.02 & 0.02 & 0.002340 & 0.006623 & 0.000049 & 1.01 & 0.83 \\ 
  retired & 0.03 & 0.03 & 0.004822 & 0.008194 & 0.000090 & 1.40 & 5.09 \\ 
  homemakr & 0.06 & 0.05 & -0.012629 & 0.008258 & 0.000228 & 1.90 & 9.95 \\ 
  disabled & 0.04 & 0.04 & -0.005893 & 0.007172 & 0.000086 & 1.03 & 3.38 \\ 
  unemploy & 0.16 & 0.13 & -0.030013 & 0.012548 & 0.001058 & 3.07 & 3.15 \\ 
  homeless & 0.01 & 0.01 & -0.000593 & 0.004002 & 0.000016 & 0.72 & 0.84 \\ 
  deg2plus & 0.82 & 0.62 & -0.206160 & 0.027626 & 0.043265 & 91.68 & 41.02 \\ 
    \hline
$\bar y$ & actual & E.est & bias & sd & mse & eff & rbias \\ 
  \hline
  degree & 7.88 & 14.24 & 6.359845 & 0.203947 & 40.489220 & 163.15 & 15.31 \\ 
  nonwhite & 0.24 & 0.30 & 0.057690 & 0.016618 & 0.003604 & 4.63 & 3.30 \\ 
  female & 0.43 & 0.46 & 0.025993 & 0.011495 & 0.000808 & 1.58 & 30.24 \\ 
  worker & 0.05 & 0.10 & 0.047894 & 0.005763 & 0.002327 & 17.17 & 7.82 \\ 
  procurer & 0.02 & 0.03 & 0.019219 & 0.003017 & 0.000378 & 12.84 & 6.47 \\ 
  client & 0.09 & 0.09 & -0.000762 & 0.007672 & 0.000059 & 0.24 & 0.17 \\ 
  dealer & 0.06 & 0.13 & 0.062733 & 0.006721 & 0.003981 & 20.09 & 6.57 \\ 
  cook & 0.01 & 0.01 & 0.001608 & 0.002173 & 0.000007 & 0.46 & 10.18 \\ 
  thief & 0.02 & 0.04 & 0.017587 & 0.004122 & 0.000326 & 6.71 & 6.23 \\ 
  retired & 0.03 & 0.03 & 0.000806 & 0.003848 & 0.000015 & 0.24 & 0.85 \\ 
  homemakr & 0.06 & 0.06 & 0.003532 & 0.006186 & 0.000051 & 0.42 & 2.78 \\ 
  disabled & 0.04 & 0.06 & 0.015092 & 0.005256 & 0.000255 & 3.04 & 8.66 \\ 
  unemploy & 0.16 & 0.26 & 0.094729 & 0.009813 & 0.009070 & 26.33 & 9.94 \\ 
  homeless & 0.01 & 0.02 & 0.004726 & 0.002590 & 0.000029 & 1.27 & 6.71 \\ 
  deg2plus & 0.82 & 0.93 & 0.103386 & 0.007816 & 0.010750 & 22.78 & 20.57 \\ 
   \hline
\end{tabular} 
\label{sbtable}
\end{table}

\begin{table}[ht]
\centering
\caption{SB Plus Table} 
\begin{tabular}{lrrrrrrr}
  \hline
name & actual & E.est & bias & sd & mse & eff & rbias \\ 
  \hline
degree & 7.88 & 7.85 & -0.033712 & 0.326760 & 0.107909 & 1.00 & 1.00 \\ 
  nonwhite & 0.24 & 0.23 & -0.014572 & 0.023778 & 0.000778 & 1.00 & 1.00 \\ 
  female & 0.43 & 0.44 & 0.008589 & 0.028383 & 0.000879 & 1.00 & 1.00 \\ 
  worker & 0.05 & 0.06 & 0.003772 & 0.010868 & 0.000132 & 1.00 & 1.00 \\ 
  procurer & 0.02 & 0.02 & 0.000492 & 0.004367 & 0.000019 & 1.00 & 1.00 \\ 
  client & 0.09 & 0.08 & -0.012597 & 0.014205 & 0.000360 & 1.00 & 1.00 \\ 
  dealer & 0.06 & 0.06 & -0.002636 & 0.009971 & 0.000106 & 1.00 & 1.00 \\ 
  cook & 0.01 & 0.01 & -0.000367 & 0.004468 & 0.000020 & 1.00 & 1.00 \\ 
  thief & 0.02 & 0.02 & -0.001448 & 0.005949 & 0.000037 & 1.00 & 1.00 \\ 
  retired & 0.03 & 0.03 & -0.002070 & 0.008906 & 0.000084 & 1.00 & 1.00 \\ 
  homemakr & 0.06 & 0.06 & -0.000533 & 0.013301 & 0.000177 & 1.00 & 1.00 \\ 
  disabled & 0.04 & 0.04 & -0.000826 & 0.010459 & 0.000110 & 1.00 & 1.00 \\ 
  unemploy & 0.16 & 0.16 & 0.001145 & 0.018468 & 0.000342 & 1.00 & 1.00 \\ 
  homeless & 0.01 & 0.01 & -0.000617 & 0.005390 & 0.000029 & 1.00 & 1.00 \\ 
  deg2plus & 0.82 & 0.81 & -0.007006 & 0.027362 & 0.000798 & 1.00 & 1.00 \\ 
    \hline
Current & actual & E.est & bias & sd & mse & eff & rbias \\ 
  \hline
  degree & 7.88 & 5.19 & -2.695417 & 0.194367 & 7.303052 & 67.68 & 79.95 \\ 
  nonwhite & 0.24 & 0.28 & 0.033953 & 0.023152 & 0.001689 & 2.17 & 2.33 \\ 
  female & 0.43 & 0.39 & -0.040886 & 0.021233 & 0.002123 & 2.41 & 4.76 \\ 
  worker & 0.05 & 0.05 & -0.002735 & 0.009258 & 0.000093 & 0.70 & 0.73 \\ 
  procurer & 0.02 & 0.01 & -0.001996 & 0.003505 & 0.000016 & 0.84 & 4.06 \\ 
  client & 0.09 & 0.15 & 0.064097 & 0.019601 & 0.004493 & 12.46 & 5.09 \\ 
  dealer & 0.06 & 0.07 & 0.008166 & 0.011082 & 0.000190 & 1.78 & 3.10 \\ 
  cook & 0.01 & 0.01 & -0.001556 & 0.002649 & 0.000009 & 0.47 & 4.24 \\ 
  thief & 0.02 & 0.02 & 0.002316 & 0.006566 & 0.000048 & 1.29 & 1.60 \\ 
  retired & 0.03 & 0.03 & 0.005184 & 0.008238 & 0.000095 & 1.13 & 2.50 \\ 
  homemakr & 0.06 & 0.05 & -0.012559 & 0.008124 & 0.000224 & 1.26 & 23.55 \\ 
  disabled & 0.04 & 0.03 & -0.006311 & 0.006973 & 0.000088 & 0.80 & 7.64 \\ 
  unemploy & 0.16 & 0.13 & -0.029973 & 0.013127 & 0.001071 & 3.13 & 26.17 \\ 
  homeless & 0.01 & 0.01 & -0.000779 & 0.003940 & 0.000016 & 0.55 & 1.26 \\ 
  deg2plus & 0.82 & 0.61 & -0.208093 & 0.026674 & 0.044014 & 55.17 & 29.70 \\ 
    \hline
$\bar y$ & actual & E.est & bias & sd & mse & eff & rbias \\ 
  \hline
  degree & 7.88 & 14.23 & 6.350125 & 0.211282 & 40.368722 & 374.10 & 188.36 \\ 
  nonwhite & 0.24 & 0.30 & 0.057748 & 0.017387 & 0.003637 & 4.68 & 3.96 \\ 
  female & 0.43 & 0.46 & 0.024594 & 0.011233 & 0.000731 & 0.83 & 2.86 \\ 
  worker & 0.05 & 0.10 & 0.047078 & 0.005580 & 0.002247 & 16.98 & 12.48 \\ 
  procurer & 0.02 & 0.03 & 0.019494 & 0.003000 & 0.000389 & 20.15 & 39.64 \\ 
  client & 0.09 & 0.09 & -0.000565 & 0.007691 & 0.000059 & 0.16 & 0.04 \\ 
  dealer & 0.06 & 0.13 & 0.062293 & 0.007300 & 0.003934 & 36.98 & 23.63 \\ 
  cook & 0.01 & 0.01 & 0.001664 & 0.002123 & 0.000007 & 0.36 & 4.54 \\ 
  thief & 0.02 & 0.04 & 0.017554 & 0.004057 & 0.000325 & 8.66 & 12.12 \\ 
  retired & 0.03 & 0.03 & 0.000898 & 0.004041 & 0.000017 & 0.21 & 0.43 \\ 
  homemakr & 0.06 & 0.06 & 0.003775 & 0.006121 & 0.000052 & 0.29 & 7.08 \\ 
  disabled & 0.04 & 0.06 & 0.014984 & 0.005194 & 0.000251 & 2.28 & 18.14 \\ 
  unemploy & 0.16 & 0.26 & 0.094894 & 0.009935 & 0.009104 & 26.59 & 82.86 \\ 
  homeless & 0.01 & 0.02 & 0.004653 & 0.002655 & 0.000029 & 0.98 & 7.54 \\ 
  deg2plus & 0.82 & 0.92 & 0.102821 & 0.007631 & 0.010630 & 13.33 & 14.68 \\ 
   \hline
\end{tabular}
\label{sbptable}
\end{table}


\begin{table}[ht]
\centering
\caption{RDS: Confidence Interval Coverage} 
\begin{tabular}{lrrr}
  \hline
name & actual & halfwidth & coverage \\ 
  \hline
degree & 7.88 & 0.55 & 0.77 \\ 
  nonwhite & 0.24 & 0.04 & 0.95 \\ 
  female & 0.43 & 0.06 & 0.98 \\ 
  worker & 0.05 & 0.02 & 0.95 \\ 
  procurer & 0.02 & 0.01 & 0.90 \\ 
  client & 0.09 & 0.03 & 0.95 \\ 
  dealer & 0.06 & 0.02 & 0.95 \\ 
  cook & 0.01 & 0.01 & 0.78 \\ 
  thief & 0.02 & 0.01 & 0.94 \\ 
  retired & 0.03 & 0.02 & 0.93 \\ 
  homemakr & 0.06 & 0.02 & 0.94 \\ 
  disabled & 0.04 & 0.02 & 0.94 \\ 
  unemploy & 0.16 & 0.03 & 0.95 \\ 
  homeless & 0.01 & 0.01 & 0.88 \\ 
  deg2plus & 0.82 & 0.07 & 1.00 \\ 
   \hline
\end{tabular}
\label{RDScoverage}
\end{table}

\begin{table}[ht]
\centering
\caption{RDS Plus: Confidence Interval Coverage} 
\begin{tabular}{lrrr}
  \hline
name & actual & halfwidth & coverage \\ 
  \hline
degree & 7.88 & 0.59 & 0.86 \\ 
  nonwhite & 0.24 & 0.05 & 0.86 \\ 
  female & 0.43 & 0.07 & 0.99 \\ 
  worker & 0.05 & 0.02 & 0.92 \\ 
  procurer & 0.02 & 0.01 & 0.69 \\ 
  client & 0.09 & 0.03 & 0.72 \\ 
  dealer & 0.06 & 0.02 & 0.78 \\ 
  cook & 0.01 & 0.01 & 0.65 \\ 
  thief & 0.02 & 0.01 & 0.65 \\ 
  retired & 0.03 & 0.02 & 0.84 \\ 
  homemakr & 0.06 & 0.03 & 0.93 \\ 
  disabled & 0.04 & 0.02 & 0.89 \\ 
  unemploy & 0.16 & 0.04 & 0.94 \\ 
  homeless & 0.01 & 0.01 & 0.67 \\ 
  deg2plus & 0.82 & 0.09 & 1.00 \\ 
   \hline
\end{tabular}
\label{RDSPcoverage}
\end{table}

\begin{table}[ht]
\centering
\caption{SB: Confidence Interval Coverage} 
\begin{tabular}{lrrr}
  \hline
name & actual & halfwidth & coverage \\ 
  \hline
degree & 7.88 & 0.56 & 0.72 \\ 
  nonwhite & 0.24 & 0.04 & 0.92 \\ 
  female & 0.43 & 0.06 & 0.99 \\ 
  worker & 0.05 & 0.02 & 0.95 \\ 
  procurer & 0.02 & 0.01 & 0.95 \\ 
  client & 0.09 & 0.03 & 0.96 \\ 
  dealer & 0.06 & 0.02 & 0.94 \\ 
  cook & 0.01 & 0.01 & 0.79 \\ 
  thief & 0.02 & 0.01 & 0.93 \\ 
  retired & 0.03 & 0.02 & 0.93 \\ 
  homemakr & 0.06 & 0.02 & 0.93 \\ 
  disabled & 0.04 & 0.02 & 0.94 \\ 
  unemploy & 0.16 & 0.03 & 0.95 \\ 
  homeless & 0.01 & 0.01 & 0.86 \\ 
  deg2plus & 0.82 & 0.07 & 1.00 \\ 
   \hline
\end{tabular}
\label{SBcoverage}
\end{table}

\begin{table}[ht]
\centering
\caption{SB Plus: Confidence Interval Coverage} 
\begin{tabular}{lrrr}
  \hline
name & actual & halfwidth & coverage \\ 
  \hline
degree & 7.88 & 0.59 & 0.92 \\ 
  nonwhite & 0.24 & 0.05 & 0.88 \\ 
  female & 0.43 & 0.07 & 0.99 \\ 
  worker & 0.05 & 0.02 & 0.94 \\ 
  procurer & 0.02 & 0.01 & 0.85 \\ 
  client & 0.09 & 0.03 & 0.79 \\ 
  dealer & 0.06 & 0.02 & 0.85 \\ 
  cook & 0.01 & 0.01 & 0.67 \\ 
  thief & 0.02 & 0.01 & 0.71 \\ 
  retired & 0.03 & 0.02 & 0.84 \\ 
  homemakr & 0.06 & 0.03 & 0.91 \\ 
  disabled & 0.04 & 0.02 & 0.89 \\ 
  unemploy & 0.16 & 0.04 & 0.94 \\ 
  homeless & 0.01 & 0.01 & 0.67 \\ 
  deg2plus & 0.82 & 0.09 & 1.00 \\ 
   \hline
\end{tabular}
\label{SBPcoverage}
\end{table}


\bibliography{fastrefs}

\begin{thebibliography}{27}

\bibitem[\protect\citeauthoryear{Baraff, McCormick and
  Raftery}{2016}]{baraff2016estimating}
\begin{barticle}[author]
\bauthor{\bsnm{Baraff},~\bfnm{Aaron~J}\binits{A.~J.}},
  \bauthor{\bsnm{McCormick},~\bfnm{Tyler~H}\binits{T.~H.}} \AND
  \bauthor{\bsnm{Raftery},~\bfnm{Adrian~E}\binits{A.~E.}}
(\byear{2016}).
\btitle{Estimating uncertainty in respondent-driven sampling using a tree
  bootstrap method}.
\bjournal{Proceedings of the National Academy of Sciences}
\bvolume{113}
\bpages{14668--14673}.
\end{barticle}
\endbibitem

\bibitem[\protect\citeauthoryear{Birnbaum and
  Sirken}{1965}]{birnbaum1965design}
\begin{barticle}[author]
\bauthor{\bsnm{Birnbaum},~\bfnm{ZW}\binits{Z.}} \AND
  \bauthor{\bsnm{Sirken},~\bfnm{Monroe~G}\binits{M.~G.}}
(\byear{1965}).
\btitle{Design of sample surveys to estimate the prevalence of rare diseases:
  three unbiased estimates, vital and health statistics, series 2}.
\bjournal{Government Printing Office, Washington, DC}.
\end{barticle}
\endbibitem

\bibitem[\protect\citeauthoryear{Brewer}{1963}]{brewer1963ratio}
\begin{barticle}[author]
\bauthor{\bsnm{Brewer},~\bfnm{KRW}\binits{K.}}
(\byear{1963}).
\btitle{Ratio estimation and finite populations: Some results deducible from
  the assumption of an underlying stochastic process}.
\bjournal{Australian Journal of Statistics}
\bvolume{5}
\bpages{93--105}.
\end{barticle}
\endbibitem

\bibitem[\protect\citeauthoryear{Brewer and Hanif}{1983}]{brewer1983sampling}
\begin{bbook}[author]
\bauthor{\bsnm{Brewer},~\bfnm{Ken~RW}\binits{K.~R.}} \AND
  \bauthor{\bsnm{Hanif},~\bfnm{Muhammad}\binits{M.}}
(\byear{1983}).
\btitle{Sampling with unequal probabilities}
\bvolume{15}.
\bpublisher{Springer Science \& Business Media}.
\end{bbook}
\endbibitem

\bibitem[\protect\citeauthoryear{Crawford}{2016}]{crawford2016graphical}
\begin{barticle}[author]
\bauthor{\bsnm{Crawford},~\bfnm{Forrest~W}\binits{F.~W.}}
(\byear{2016}).
\btitle{The graphical structure of respondent-driven sampling}.
\bjournal{Sociological methodology}
\bvolume{46}
\bpages{187--211}.
\end{barticle}
\endbibitem

\bibitem[\protect\citeauthoryear{Crawford, Wu and
  Heimer}{2018}]{crawford2018hidden}
\begin{barticle}[author]
\bauthor{\bsnm{Crawford},~\bfnm{Forrest~W}\binits{F.~W.}},
  \bauthor{\bsnm{Wu},~\bfnm{Jiacheng}\binits{J.}} \AND
  \bauthor{\bsnm{Heimer},~\bfnm{Robert}\binits{R.}}
(\byear{2018}).
\btitle{Hidden population size estimation from respondent-driven sampling: a
  network approach}.
\bjournal{Journal of the American Statistical Association}
\bvolume{113}
\bpages{755--766}.
\end{barticle}
\endbibitem

\bibitem[\protect\citeauthoryear{Fellows}{2018}]{fellows2018respondent}
\begin{barticle}[author]
\bauthor{\bsnm{Fellows},~\bfnm{Ian~E}\binits{I.~E.}}
(\byear{2018}).
\btitle{Respondent-driven sampling and the homophily configuration graph}.
\bjournal{Statistics in medicine}.
\end{barticle}
\endbibitem

\bibitem[\protect\citeauthoryear{Frank}{1977}]{frank1977survey}
\begin{barticle}[author]
\bauthor{\bsnm{Frank},~\bfnm{Ove}\binits{O.}}
(\byear{1977}).
\btitle{Survey sampling in graphs}.
\bjournal{Journal of Statistical Planning and Inference}
\bvolume{1}
\bpages{235--264}.
\end{barticle}
\endbibitem

\bibitem[\protect\citeauthoryear{Frank and
  Snijders}{1994}]{frank1994estimating}
\begin{barticle}[author]
\bauthor{\bsnm{Frank},~\bfnm{Ove}\binits{O.}} \AND
  \bauthor{\bsnm{Snijders},~\bfnm{Tom}\binits{T.}}
(\byear{1994}).
\btitle{Estimating the size of hidden populations using snowball sampling}.
\bjournal{Journal of Official Statistics}
\bvolume{10}
\bpages{53--53}.
\end{barticle}
\endbibitem

\bibitem[\protect\citeauthoryear{Gile}{2011}]{gile2011improved}
\begin{barticle}[author]
\bauthor{\bsnm{Gile},~\bfnm{Krista~J}\binits{K.~J.}}
(\byear{2011}).
\btitle{Improved inference for respondent-driven sampling data with application
  to HIV prevalence estimation}.
\bjournal{Journal of the American Statistical Association}
\bvolume{106}
\bpages{135--146}.
\end{barticle}
\endbibitem

\bibitem[\protect\citeauthoryear{Gile and Handcock}{2010}]{gile20107}
\begin{barticle}[author]
\bauthor{\bsnm{Gile},~\bfnm{Krista~J}\binits{K.~J.}} \AND
  \bauthor{\bsnm{Handcock},~\bfnm{Mark~S}\binits{M.~S.}}
(\byear{2010}).
\btitle{7. Respondent-Driven Sampling: An Assessment of Current Methodology}.
\bjournal{Sociological methodology}
\bvolume{40}
\bpages{285--327}.
\end{barticle}
\endbibitem

\bibitem[\protect\citeauthoryear{Goel and Salganik}{2010}]{goel2010assessing}
\begin{barticle}[author]
\bauthor{\bsnm{Goel},~\bfnm{Sharad}\binits{S.}} \AND
  \bauthor{\bsnm{Salganik},~\bfnm{Matthew~J}\binits{M.~J.}}
(\byear{2010}).
\btitle{Assessing respondent-driven sampling}.
\bjournal{Proceedings of the National Academy of Sciences}
\bvolume{107}
\bpages{6743--6747}.
\end{barticle}
\endbibitem

\bibitem[\protect\citeauthoryear{Heckathorn}{1997}]{heckathorn1997respondent}
\begin{barticle}[author]
\bauthor{\bsnm{Heckathorn},~\bfnm{Douglas~D}\binits{D.~D.}}
(\byear{1997}).
\btitle{Respondent-driven sampling: a new approach to the study of hidden
  populations}.
\bjournal{Social problems}
\bvolume{44}
\bpages{174--199}.
\end{barticle}
\endbibitem

\bibitem[\protect\citeauthoryear{Heckathorn}{2007}]{heckathorn20076}
\begin{barticle}[author]
\bauthor{\bsnm{Heckathorn},~\bfnm{Douglas~D}\binits{D.~D.}}
(\byear{2007}).
\btitle{6. Extensions of Respondent-Driven Sampling: Analyzing Continuous
  Variables and Controlling for Differential Recruitment}.
\bjournal{Sociological Methodology}
\bvolume{37}
\bpages{151--208}.
\end{barticle}
\endbibitem

\bibitem[\protect\citeauthoryear{Potterat, Rothenberg and
  Muth}{1999}]{potterat1999network}
\begin{barticle}[author]
\bauthor{\bsnm{Potterat},~\bfnm{John~J}\binits{J.~J.}},
  \bauthor{\bsnm{Rothenberg},~\bfnm{Richard~B}\binits{R.~B.}} \AND
  \bauthor{\bsnm{Muth},~\bfnm{Steven~Q}\binits{S.~Q.}}
(\byear{1999}).
\btitle{Network structural dynamics and infectious disease propagation}.
\bjournal{International journal of STD \& AIDS}
\bvolume{10}
\bpages{182--185}.
\end{barticle}
\endbibitem

\bibitem[\protect\citeauthoryear{Salganik}{2006}]{salganik2006variance}
\begin{barticle}[author]
\bauthor{\bsnm{Salganik},~\bfnm{Matthew~J}\binits{M.~J.}}
(\byear{2006}).
\btitle{Variance estimation, design effects, and sample size calculations for
  respondent-driven sampling}.
\bjournal{Journal of Urban Health}
\bvolume{83}
\bpages{98}.
\end{barticle}
\endbibitem

\bibitem[\protect\citeauthoryear{Salganik and
  Heckathorn}{2004}]{salganik2004sampling}
\begin{barticle}[author]
\bauthor{\bsnm{Salganik},~\bfnm{Matthew~J}\binits{M.~J.}} \AND
  \bauthor{\bsnm{Heckathorn},~\bfnm{Douglas~D}\binits{D.~D.}}
(\byear{2004}).
\btitle{Sampling and estimation in hidden populations using respondent-driven
  sampling}.
\bjournal{Sociological methodology}
\bvolume{34}
\bpages{193--240}.
\end{barticle}
\endbibitem

\bibitem[\protect\citeauthoryear{S{\"a}rndal, Swensson and
  Wretman}{2003}]{sarndal2003model}
\begin{bbook}[author]
\bauthor{\bsnm{S{\"a}rndal},~\bfnm{Carl-Erik}\binits{C.-E.}},
  \bauthor{\bsnm{Swensson},~\bfnm{Bengt}\binits{B.}} \AND
  \bauthor{\bsnm{Wretman},~\bfnm{Jan}\binits{J.}}
(\byear{2003}).
\btitle{Model assisted survey sampling}.
\bpublisher{Springer Science \& Business Media}.
\end{bbook}
\endbibitem

\bibitem[\protect\citeauthoryear{Spiller et~al.}{2017}]{spiller2017evaluating}
\begin{barticle}[author]
\bauthor{\bsnm{Spiller},~\bfnm{Michael~W}\binits{M.~W.}},
  \bauthor{\bsnm{Gile},~\bfnm{Krista~J}\binits{K.~J.}},
  \bauthor{\bsnm{Handcock},~\bfnm{Mark~S}\binits{M.~S.}},
  \bauthor{\bsnm{Mar},~\bfnm{Corinne~M}\binits{C.~M.}} \AND
  \bauthor{\bsnm{Wejnert},~\bfnm{Cyprian}\binits{C.}}
(\byear{2017}).
\btitle{Evaluating variance estimators for respondent-driven sampling}.
\bjournal{Journal of survey statistics and methodology}
\bvolume{6}
\bpages{23--45}.
\end{barticle}
\endbibitem

\bibitem[\protect\citeauthoryear{Spreen}{1992}]{spreen1992rare}
\begin{barticle}[author]
\bauthor{\bsnm{Spreen},~\bfnm{Marinus}\binits{M.}}
(\byear{1992}).
\btitle{Rare populations, hidden populations, and link-tracing designs: What
  and why?}
\bjournal{Bulletin of Sociological Methodology/Bulletin de Methodologie
  Sociologique}
\bvolume{36}
\bpages{34--58}.
\end{barticle}
\endbibitem

\bibitem[\protect\citeauthoryear{Thompson}{2006}]{thompson2006aws}
\begin{barticle}[author]
\bauthor{\bsnm{Thompson},~\bfnm{Steven~K.}\binits{S.~K.}}
(\byear{2006}).
\btitle{Adaptive Web Sampling}.
\bjournal{Biometrics}
\bvolume{62}
\bpages{1224--1234}.
\bdoi{10.1111/j.1541-0420.2006.00576.x}
\end{barticle}
\endbibitem

\bibitem[\protect\citeauthoryear{Thompson}{2015}]{thompson2015fast}
\begin{barticle}[author]
\bauthor{\bsnm{Thompson},~\bfnm{Steven~K}\binits{S.~K.}}
(\byear{2015}).
\btitle{Fast Moving Sampling Designs in Temporal Networks}.
\bjournal{arXiv preprint arXiv:1511.09149}.
\end{barticle}
\endbibitem

\bibitem[\protect\citeauthoryear{Thompson}{2017}]{thompson2017adaptive}
\begin{barticle}[author]
\bauthor{\bsnm{Thompson},~\bfnm{Steven~K}\binits{S.~K.}}
(\byear{2017}).
\btitle{Adaptive and Network Sampling for Inference and Interventions in
  Changing Populations}.
\bjournal{Journal of Survey Statistics and Methodology}
\bvolume{5}
\bpages{1--21}.
\end{barticle}
\endbibitem

\bibitem[\protect\citeauthoryear{Thompson}{2018}]{thompson2018simple}
\begin{barticle}[author]
\bauthor{\bsnm{Thompson},~\bfnm{Steve}\binits{S.}}
(\byear{2018}).
\btitle{Simple estimators for network sampling}.
\bjournal{arXiv preprint arXiv:1804.00808}.
\end{barticle}
\endbibitem

\bibitem[\protect\citeauthoryear{Thompson and
  Collins}{2002}]{thompson2002adaptive}
\begin{barticle}[author]
\bauthor{\bsnm{Thompson},~\bfnm{Steven~K}\binits{S.~K.}} \AND
  \bauthor{\bsnm{Collins},~\bfnm{Linda~M}\binits{L.~M.}}
(\byear{2002}).
\btitle{Adaptive sampling in research on risk-related behaviors}.
\bjournal{Drug and Alcohol Dependence}
\bvolume{68}
\bpages{57--67}.
\end{barticle}
\endbibitem

\bibitem[\protect\citeauthoryear{Volz and
  Heckathorn}{2008}]{volz2008probability}
\begin{barticle}[author]
\bauthor{\bsnm{Volz},~\bfnm{Erik}\binits{E.}} \AND
  \bauthor{\bsnm{Heckathorn},~\bfnm{Douglas~D}\binits{D.~D.}}
(\byear{2008}).
\btitle{Probability based estimation theory for respondent driven sampling}.
\bjournal{Journal of official statistics}
\bvolume{24}
\bpages{79}.
\end{barticle}
\endbibitem

\bibitem[\protect\citeauthoryear{Young, Rudolph and
  Havens}{2018}]{young2018network}
\begin{barticle}[author]
\bauthor{\bsnm{Young},~\bfnm{April~M}\binits{A.~M.}},
  \bauthor{\bsnm{Rudolph},~\bfnm{Abby~E}\binits{A.~E.}} \AND
  \bauthor{\bsnm{Havens},~\bfnm{Jennifer~R}\binits{J.~R.}}
(\byear{2018}).
\btitle{Network-based research on rural opioid use: an overview of methods and
  lessons learned}.
\bjournal{Current HIV/AIDS Reports}
\bpages{1--7}.
\end{barticle}
\endbibitem

\end{thebibliography}

\end{document}